\documentclass[aip,jcp,graphicx,groupedaddress,reprint]{revtex4-1}
\usepackage{amsmath}
\usepackage{amssymb}
\usepackage{graphicx}
\usepackage{hyperref}
\usepackage{tensor}
\usepackage{txfontsb} %most up-to-date times font supported by the arXiv
\usepackage{multirow}
\DeclareMathAlphabet{\mathcal}{OMS}{cmsy}{m}{n} %Keep the nice script-O
\usepackage{algpseudocode} %for algorithm blocks ...
\usepackage{etoolbox}
\usepackage{newfloat}

\AtEndEnvironment{algorithm}{\vskip-6pt\noindent\hrulefill}
\AtBeginEnvironment{algorithmic}{\vskip-7pt\noindent\hrulefill}
\DeclareFloatingEnvironment[
    fileext=loa,
    listname=List of Algorithms,
    name=ALGORITHM,
    placement=tbhp,
]{algorithm}
\usepackage{rotating} %for rotated tables ...
\hypersetup{
    colorlinks=true,
    linkcolor=blue,
    citecolor=blue,
    filecolor=blue,
    urlcolor=blue,
}

\begin{document}

\title{Cubic-scaling algorithm and self-consistent field
 for the random-phase approximation with second-order screened exchange}

\author{Jonathan E. Moussa}
\email{godotalgorithm@gmail.com}
\affiliation{Sandia National Laboratories, Albuquerque, NM 87185, USA}

\begin{abstract}  \begin{minipage}{0.79\textwidth}
\ \\

The random-phase approximation with second-order screened exchange (RPA+SOSEX) is a model
 of electron correlation energy with two caveats:
 its accuracy depends on an arbitrary choice of mean field, and it scales as $\mathcal{O}(n^5)$ operations
 and $\mathcal{O}(n^3)$ memory for $n$ electrons.
We derive a new algorithm that reduces its scaling to $\mathcal{O}(n^3)$ operations 
 and $\mathcal{O}(n^2)$ memory using controlled approximations and a new self-consistent field
 that approximates Brueckner coupled-cluster doubles (BCCD) theory with RPA+SOSEX,
 referred to as Brueckner RPA (BRPA) theory.
The algorithm comparably reduces the scaling of second-order M{\o}ller-Plesset (MP2) perturbation theory
 with smaller cost prefactors than RPA+SOSEX.
Within a semiempirical model, we study H$_2$ dissociation to test accuracy and H$_n$ rings to verify scaling.
%\textit{\copyright \ 2013 AIP Publishing LLC.} [http://dx.doi.org/10.1063/1.477wxyz] %for spacing estimates ...
\end{minipage}\end{abstract}

%DOESN'T APPEAR IN JCP PAPERS, JUST ON THE WEBSITE
%02.60.Dc, Numerical linear algebra
%31.15.bw, Coupled-cluster theory
%31.15.xr, Self-consistent-field methods
%71.15.-m, Methods of electronic structure calculations
%\pacs{02.60.Dc,31.15.bw,31.15.xr,71.15.-m}

\maketitle

%S1
\section{Introduction}
\vspace{-7pt}

%P1.1 - DFT & CC are the two main many-electron ground state methods
Density functional theory (DFT) and coupled-cluster (CC) theory are the two dominant paradigms
 for the computation of many-electron ground states with complementary capabilities.
Both theories are built upon the  independent-orbital and self-consistent field (SCF) structure of Hartree-Fock (HF) theory \cite{QCtextbook}.
DFT proves the existence of an exact density functional \cite{HohenbergKohn} for which
 there are approximations \cite{KohnSham,LDA} enabling routine simulation of thousands of electrons \cite{bigDFT}
 that scale as $\mathcal{O}(n^3)$ operations and $\mathcal{O}(n^2)$ memory for $n$ electrons.
CC theory is a systematically improvable hierarchy of methods \cite{CCreview} indexed by $p \ge 2$
 that scale as $\mathcal{O}(n^{2p+2})$ operations and $\mathcal{O}(n^{2p})$ memory.
In practice, DFT is limited by accuracy and CC theory is limited by cost.

%P1.2 - Convergence of DFT & CC to a common point: RPA
The random phase approximation (RPA) is a natural point of convergence
 in the ongoing development of CC theory and DFT.
RPA emerges from truncated versions of CC methods as
 reduced-complexity methods that retain a significant fraction of the CC correlation energy \cite{RPAfromCC}.
RPA also occurs when DFT is used to approximate the polarization function in the
 adiabatic-connection fluctuation-dissipation formula for the correlation energy \cite{RPAfromACFD}.
This formula might become a part of more accurate density functionals,
 but it has a higher cost and complexity \cite{oldfastRPA} than conventional density functionals.
In both cases, a balance of cost and accuracy is being sought with RPA.

%P1.3 - RPA+SOSEX as a minimal nonempirical RPA model
The electron correlation energy model that we consider in this paper is the RPA correlation energy
 plus the second-order screened exchange (SOSEX) energy \cite{VASP_SOSEX}.
RPA+SOSEX is exact for one electron and to second order in perturbation theory.
It agrees with quantum Monte Carlo benchmarks of the uniform electron gas \cite{QMC_egas}
 to within 0.002 Ha/electron \cite{Freeman_SOSEX}.
Benchmarks on inhomogeneous systems show mean absolute errors of 
 0.002 Ha/atom for cohesive energies of solids in a 5-solid test set \cite{VASP_SOSEX},
 0.008 Ha/molecule for atomization energies of molecules in the G2-1 test set \cite{bench_SOSEX},
 and 0.010 Ha/atom for correlation energies of first and second row atoms \cite{bench_SOSEX}.
These atomic and molecular RPA+SOSEX results approach but fail to surpass the accuracy
 of second-order M{\o}ller-Plesset (MP2) perturbation theory \cite{RPA3rd,SOSEX_implementation}
 and the B3LYP density functional \cite{SE_SOSEX}.
Also, any RPA+SOSEX energy depends on a mean field choice and is not unique \cite{VASP_SOSEX}.

%P1.4 - Basic summary of goals
The goal of this paper is to further develop RPA+SOSEX as a compromise
 between the cost of DFT and the accuracy of CC theory.
Known algorithms for RPA+SOSEX \cite{VASP_SOSEX} and similar models \cite{SOSEX_implementation}
 require $\mathcal{O}(n^5)$ operations and either $\mathcal{O}(n^3)$ or $\mathcal{O}(n^4)$ memory.
By utilizing an auxiliary basis set \cite{auxiliary}, fast interaction kernel summation \cite{fast_kernel_sum},
 and a low-rank approximation of energy denominators \cite{Laplace},
 we design a new algorithm to reduce the cost to $\mathcal{O}(n^3)$ operations and $\mathcal{O}(n^2)$ memory.
We define a precise RPA+SOSEX total energy with a unique choice of mean field based on Brueckner orbitals \cite{Brueckner_orbitals}
 that is designed to approximate Brueckner coupled-cluster doubles (BCCD) theory \cite{BCCD}.
This is referred to as Brueckner RPA (BRPA) theory.

%P1.5 - No application space for RPA+SOSEX separate from DFT or CC
It is important to be pragmatic about the short-term value of BRPA theory and a fast RPA+SOSEX algorithm.
Despite its construction, BRPA is not a good approximation of BCCD.
The approximations used to retain the RPA+SOSEX form are too crude.
For a given basis, a fast RPA+SOSEX calculation
 will have a large cost relative to conventional DFT.
With the continued reduction of average errors in density functionals \cite{DFTaccuracy}
 and a correlation of DFT and RPA error outliers \cite{DFTerrors},
 it is unclear whether the cost is warranted without additional research.

%P1.6 - Outline of the paper
The paper proceeds as follows.
BRPA theory is derived in Sec. \ref{BRPAtheory} as a truncation of BCCD theory.
The SCF structure of BRPA theory is emphasized.
The three main components of a fast RPA+SOSEX algorithm
 are introduced in Sec. \ref{fast_components}.
This includes a nonstandard choice of primary and auxiliary basis,
 which simplifies the cost accounting and algorithm design.
In Sec. \ref{RPAstructure}, the tensor structure in existing RPA
 algorithms \cite{oldfastRPA,SOSEX_implementation} is reviewed and extended to RPA+SOSEX.
The novel structure of RPA+SOSEX theory enables a significant reduction in the number of its variables.
Fast and conventional RPA+SOSEX algorithms are designed in Sec. \ref{fast_design} as pseudocode.
A detailed leading-order cost analysis is given instead of a crude scaling analysis.
We study applications to a semiempirical H$_n$ model in Sec. \ref{applications}.
The accuracy of BRPA theory is tested on H$_2$, and the scaling of RPA+SOSEX algorithms
 is confirmed with H$_n$ calculations for large $n$.
The implications for future work are discussed in Sec. \ref{discussion}.
This includes implementation in established electronic structure codes,
 basis set convergence problems, and further development of correlation models.
We conclude in Sec. \ref{conclusions} with a brief consideration of RPA as a distinct electronic structure paradigm.

%artificial first-page termination
\newpage

%S2
\section{Brueckner RPA (BRPA) theory\label{BRPAtheory}}
\vspace{-7pt}

%P2.1 - Second-quantized Hamiltonian
The common origin of all post-HF methods is the many-electron Hamiltonian in second quantization notation \cite{QCtextbook},
\begin{equation}
 \hat{H} = E_0 + h_q^p \hat{c}_p^\dag \hat{c}_q + \tfrac{1}{4} \widetilde{V}_{rs}^{pq} \hat{c}_p^\dag \hat{c}_q^\dag \hat{c}_s \hat{c}_r,
\end{equation}
 where $\hat{c}_p$ are the fermion lowering operators of spin-orbitals.
Spin-orbital indices are labeled by $\{ p,q,r,s \}$.
Any tensor or product of tensors with repeated indices has an implicit sum over those indices
 unless they appear unrepeated in a tensor or product of tensors
 or in an explicit sum in the same equation.
This notation is related to the standard bracket notation \cite{QCtextbook} as
\begin{align}
 h_q^p &= \langle p | h | q \rangle = \int d\mathbf{x} d\mathbf{y} \phi_p^*(\mathbf{x}) h(\mathbf{x},\mathbf{y}) \phi_q(\mathbf{y}) , \notag \\
 V_{rs}^{pq} &= \langle p q | r s \rangle = \int d\mathbf{x} d\mathbf{y} \phi_p^*(\mathbf{x}) \phi_q^*(\mathbf{y}) V(\mathbf{x},\mathbf{y}) \phi_r(\mathbf{x}) \phi_s(\mathbf{y}) , \notag \\
 \widetilde{V}_{rs}^{pq} &=  \langle p q || r s \rangle = \langle p q | r s \rangle - \langle p q | s r \rangle ,
\end{align}
 where $\mathbf{x}$ and $\mathbf{y}$ are spin-space coordinates,
 $h(\mathbf{x},\mathbf{y})$ is the single-electron Hamiltonian,
 $V(\mathbf{x},\mathbf{y})$ is the inter-electron interaction,
 $\phi_p(\mathbf{x})$ are the spin-orbital wavefunctions,
 and $\widetilde{X}_{rs}^{pq} = X_{rs}^{pq} - X_{sr}^{pq}$ is a tensor antisymmetrization operation.
The operators have symmetries $h(\mathbf{x},\mathbf{y}) = h(\mathbf{y},\mathbf{x})^*$ and $V(\mathbf{x},\mathbf{y}) = V(\mathbf{y},\mathbf{x})$.

%P2.2 - Occupied/unoccupied & the Slater determinant
We define a reference Slater determinant, $|\Phi\rangle$, by splitting the set of orbitals
 into virtual orbitals labeled by $\{ a,b,c,d \}$ and $n$ occupied orbitals labeled by $\{ i,j,k,l \}$.
$|\Phi\rangle$ is defined by
\begin{equation} \label{phi_define}
 \hat{c}_i^\dag |\Phi\rangle = 0, \ \ \ \hat{c}_a |\Phi\rangle = 0,  \ \ \ \langle \Phi | \Phi \rangle = 1,
\end{equation}
 uniquely up to a global phase.
When combined with fermion anticommutation,
 $\hat{c}_p^\dag \hat{c}_q = \delta_q^p - \hat{c}_q \hat{c}_p^\dag$
 and $\hat{c}_p \hat{c}_q = - \hat{c}_q \hat{c}_p$, Eq. (\ref{phi_define})
 is sufficient to calculate all expectation values of $|\Phi\rangle$.

%P2.3 - HF theory & SCF structure
HF theory posits the minimization of $E_{\mathrm{HF}} = \langle \Phi | \hat{H} | \Phi \rangle$
 and diagonalization of particle and hole subspaces, $\langle \Phi | \hat{c}_a \hat{H} \hat{c}_b^\dag | \Phi \rangle$
 and $\langle \Phi | \hat{c}_i^\dag \hat{H} \hat{c}_j | \Phi \rangle$.
For local minima in $E_{\mathrm{HF}}$, this is equivalent to
 diagonalization of a Fock matrix,
\begin{equation} \label{Fock}
 f_q^{p0} = h_q^p + \widetilde{V}_{iq}^{ip} = \epsilon_p \delta_q^p ,
\end{equation}
 or equivalently a Fock operator in spin-space,
\begin{align} \label{SCF_space}
  f^0(\mathbf{x},\mathbf{y}) &=  h(\mathbf{x},\mathbf{y}) + v(\mathbf{x}) \delta(\mathbf{x}-\mathbf{y}) - \rho(\mathbf{x},\mathbf{y}) V(\mathbf{x},\mathbf{y}), \notag \\
   \rho(\mathbf{x},\mathbf{y}) &= \phi_i(\mathbf{x}) \phi_i^*(\mathbf{y}), \ \ \
    v(\mathbf{x}) =  \int d\mathbf{y} V(\mathbf{x},\mathbf{y}) \rho(\mathbf{y},\mathbf{y}), \notag \\
    & \ \ \ \ \int d\mathbf{y} f^0(\mathbf{x},\mathbf{y}) \phi_p(\mathbf{y}) = \epsilon_p \phi_p(\mathbf{x}) ,
\end{align}
 with the canonical SCF structure between $\phi_p(\mathbf{x})$ and $f^0(\mathbf{x},\mathbf{y})$.
The total energy consistent with the Fock matrix is
\begin{subequations} \begin{align}
 E_{\mathrm{HF}} &= E_0 + h_i^i + \tfrac{1}{2} \widetilde{V}_{ij}^{ij} \\
     &= E_0 + \tfrac{1}{2} ( h_i^i + f_i^{i0}). \label{HF_energy}
\end{align} \end{subequations}
The orbital energies are approximate excitation energies,
\begin{align} \label{Koopmans}
 \epsilon_a  &= \langle \Phi | \hat{c}_a \hat{H} \hat{c}_a^\dag | \Phi \rangle - E_{\mathrm{HF}}, \notag \\
  \epsilon_i &= E_{\mathrm{HF}} - \langle \Phi | \hat{c}_i^\dag \hat{H} \hat{c}_i | \Phi \rangle,
\end{align}
 which is known as Koopmans' theorem \cite{Koopmans}.

%S2.1
\subsection{Brueckner coupled-cluster doubles (BCCD) theory}
\vspace{-7pt}

%P2.4 - Basic structure of CC, CCSD, & BCCD
The basic structure of CC theory is to solve $\hat{H} |\Psi\rangle = E |\Psi\rangle$
 approximately as a projected eigenvalue problem \cite{CCreview}.
It defines a cluster operator, $\hat{T}$, from the linear span of a set of operators, $\mathcal{S}$,
 omitting identity, $\hat{I}$, which only alters the normalization,
\begin{align}
 \langle \Phi | \hat{X}^\dag & \exp(-\hat{T}) \hat{H} |\Psi\rangle = E \langle \Phi | \hat{X}^\dag \exp(-\hat{T}) |\Psi\rangle , \ \ \ 
  \hat{X} \in \mathcal{S}, \notag \\
  |\Psi\rangle &= \exp(\hat{T}) |\Phi\rangle, \ \ \ \hat{T} \in \mathrm{span}(\mathcal{S} \setminus \{ \hat{I} \}) .
\end{align}
At the singles-and-doubles level of theory (CCSD), $| \Phi \rangle$ is the HF ground state and
$\mathcal{S} = \{ \hat{I}, \hat{c}_a^\dag \hat{c}_i , \hat{c}_a^\dag \hat{c}_b^\dag \hat{c}_j \hat{c}_i \}$.
In its Brueckner variant (BCCD), $| \Phi \rangle$ is varied
 and $\hat{T} \in \mathrm{span}(\mathcal{S} \setminus \{ \hat{I}, \hat{c}_a^\dag \hat{c}_i \} )$.

% with $| \Phi \rangle$ usually set to the HF ground state.
%At the singles-and-doubles level of CC theory (CCSD),
%$\mathcal{S} = \{ 1, \hat{c}_a^\dag \hat{c}_i , \hat{c}_a^\dag \hat{c}_b^\dag \hat{c}_j \hat{c}_i \}$.
%In the Brueckner variant of this theory (BCCD), $| \Phi \rangle$ is allowed to vary
% and $\hat{T}$ is restricted to $\mathrm{span}(\mathcal{S} \setminus \{ 1, \hat{c}_a^\dag \hat{c}_i \} )$.
%In either case, there is an equal number of variables and equations.

%P2.5 - T antisymmetry & Brueckner mean field
We use nonstandard notation to simplify the tensor form of the BCCD equations.
The first is $\hat{T} = \tfrac{1}{2} T_{ij}^{ab} \hat{c}_a^\dag \hat{c}_b^\dag \hat{c}_j \hat{c}_i $
 with partial symmetry constraints: $T_{ij}^{ab} = T_{ji}^{ba}$ but not $T_{ij}^{ab} = -T_{ji}^{ab}$.
The second is a non-Hermitian Brueckner matrix \cite{BCCD},
\begin{equation}
  b_q^p = f_q^{p0} + (\delta_a^p \delta_q^i f_b^{j0} +  \tfrac{1}{2} \delta_q^i \widetilde{V}_{ab}^{pj}
 - \tfrac{1}{2} \delta_a^p \widetilde{V}_{qb}^{ij}) \widetilde{T}_{ij}^{ab} .
\end{equation}
This reduces the total energy to a form similar to Eq. (\ref{HF_energy}),
\begin{subequations}\begin{align}
 E &= E_0 + h_i^i + \tfrac{1}{2} \widetilde{V}_{ij}^{ij} + \tfrac{1}{4} \widetilde{V}_{ab}^{ij} \widetilde{T}_{ij}^{ab} \\
 &= E_0 + \tfrac{1}{2} (h_i^i +  b_i^i) . \label{CC_energy}
\end{align}\end{subequations}
Also, the single-excitation equations simplify to $b_i^a = 0$.

%P2.6 - BCCD double-excitation equations
The most complicated part of BCCD theory is the double-excitation equations.
We separate out a ``ring'' tensor,
\begin{equation}
 R_{ij}^{ab} = \widetilde{V}_{ic}^{ak} \widetilde{T}_{kj}^{cb} + \widetilde{T}_{ik}^{ac} \widetilde{V}_{cj}^{kb}
 + \widetilde{T}_{ik}^{ac} \widetilde{V}_{cd}^{kl} \widetilde{T}_{lj}^{db} ,
\end{equation}
 which simplifies the remaining equations to
\begin{align}\label{BCCD_double}
 \widetilde{V}_{ij}^{ab} &+ b_c^a \widetilde{T}_{ij}^{cb} - b_i^k \widetilde{T}_{kj}^{ab}
 + b_c^b\widetilde{T}_{ij}^{ac} - b_j^k \widetilde{T}_{ik}^{ab} + \widetilde{R}_{ij}^{ab} \notag \\
 &+ \tfrac{1}{2} \widetilde{V}_{cd}^{ab} \widetilde{T}_{ij}^{cd}
 + \tfrac{1}{2} \widetilde{T}_{kl}^{ab} \widetilde{V}_{ij}^{kl}
 + \tfrac{1}{4} \widetilde{T}_{kl}^{ab} \widetilde{V}_{cd}^{kl} \widetilde{T}_{ij}^{cd} = 0.
\end{align}
Only $\widetilde{T}_{ij}^{ab}$ appears in the BCCD tensor equations.
Changes in $T_{ij}^{ab}$ that do not alter $\widetilde{T}_{ij}^{ab}$ are a redundancy in the theory.

%S2.2
\vspace{-2pt}
\subsection{BRPA as a truncation of BCCD}
\vspace{-7pt}

%P2.7 - Basic philosophy of approximation
We construct BRPA theory by truncating the BCCD tensor equations to
 extract the RPA+SOSEX correlation model.
We crudely attempt to minimize errors by minimizing the number of truncations.
Only Eq. (\ref{BCCD_double}) is truncated by (1) removing the ``ladder'' terms on the second line of the equation,
 (2) reducing the ring tensor to a ``direct-ring'' tensor, $ \widetilde{R}_{ij}^{ab} \Rightarrow  \widetilde{D}_{ij}^{ab}$, for
\begin{equation} \label{direct_ring}
 D_{ij}^{ab} = V_{ic}^{ak} T_{kj}^{cb} + T_{ik}^{ac} V_{cj}^{kb}
 + T_{ik}^{ac} V_{cd}^{kl} T_{lj}^{db} ,
\end{equation}
 and (3) reducing $b_q^p$ to its Hermitian part, $\tilde{b}_q^p = \tfrac{1}{2} (b_q^p + b_p^{q*})$,
 to guarantee orthogonal orbitals with real energies.
The resulting equations depend on $T_{ij}^{ab}$ but only define $\widetilde{T}_{ij}^{ab}$.
To define $T_{ij}^{ab}$, we expand the set of equations by removing all `$\widetilde{ \ \ \ }$',
\begin{equation}\label{BRPA_double}
 V_{ij}^{ab} + \tilde{b}_c^a T_{ij}^{cb} - \tilde{b}_i^k T_{kj}^{ab}
 + \tilde{b}_c^b T_{ij}^{ac} - \tilde{b}_j^k T_{ik}^{ab} + D_{ij}^{ab} = 0.
\end{equation}
These are known as the RPA Riccati equations \cite{RPAfromCC}.
They have a unique, positive-definite ``stabilizing'' solution \cite{Riccati} that evolves
 over $V_{ij}^{ab} \Rightarrow \lambda V_{ij}^{ab}$ from the unique $\lambda \rightarrow 0$ solution to $\lambda=1$.

%S2.3
\subsection{Self-consistent field structure of BRPA\label{SCF_BRPA}}
\vspace{-7pt}

%P2.8 - Group BRPA equations into SCF structure
In conjunction with Eqs. (\ref{direct_ring}) and (\ref{BRPA_double}),
 it is convenient to describe BRPA theory with SCF structure.
We expand Eq. (\ref{Fock}) to the diagonalization of a generalized Fock matrix,
\begin{equation} \label{general_Fock}
 f_q^p = f_q^{p0} + \Sigma_q^p = \epsilon_p \delta_q^p,
\end{equation}
 with a static and Hermitian self-energy matrix, $\Sigma_q^p$, in analogy to many-body Green's function theory \cite{QCtextbook}.
With the choice
\begin{equation}
 f_b^a = \tilde{b}_b^a, \ \ \ f_j^i = \tilde{b}_j^i, \ \ \ f_i^a = b_i^a, \ \ \ f_a^i = b_i^{a*} ,
\end{equation}
Eq. (\ref{general_Fock}) contains $b_i^a = 0$, Eqs. (\ref{HF_energy}) and (\ref{CC_energy}) are analogous,
 and $f_q^p$ diagonalization enables a rearrangement of Eq. (\ref{BRPA_double}),
\begin{equation} \label{diag_Riccati}
 T_{ij}^{ab} = - \frac{(\delta_p^a \delta_i^r + T_{ik}^{ac} \delta_p^k \delta_c^r )V_{rs}^{pq}
 (\delta_q^l \delta_d^s T_{lj}^{db} + \delta_q^b \delta_j^s)}{\epsilon_a - \epsilon_i + \epsilon_b - \epsilon_j}.
\end{equation}
It is convenient to write $\Sigma_q^p$ using a non-Hermitian matrix,
\begin{align} \label{sigma}
 \Sigma_q^p &= \tfrac{1}{2} (\sigma_q^p + \sigma_p^{q*} +
  \delta_a^p \delta_q^i \sigma_i^a + \delta_p^i \delta_a^q \sigma_i^{a*}) , \notag \\
 \sigma_q^p &= (\delta_a^p \delta_q^i h_b^j +  \tfrac{1}{2} \delta_q^i \widetilde{V}_{ab}^{pj}
 - \tfrac{1}{2} \delta_a^p \widetilde{V}_{qb}^{ij}) \widetilde{T}_{ij}^{ab},
\end{align}
 whereas $\Sigma_q^p \Rightarrow 0$ in HF theory.

%P2.9 - Generalized SCF cycle
A generalized SCF cycle is used for BRPA calculations.
For a given $\{\epsilon_p, \phi_p(\mathbf{x})\}$, we solve
 Eq. (\ref{diag_Riccati}) iteratively for $T_{ij}^{ab}$ as an inner SCF cycle.
We then calculate $\sigma_q^p$ from $T_{ij}^{ab}$, which is used to recalculate
 $\{\epsilon_p, \phi_p(\mathbf{x})\}$ in the outer SCF cycle.
This is appropriate for reducing the number of $\sigma_q^p$ calculations
 when there is a large relative cost to calculate $\sigma_q^p$ over $T_{ij}^{ab}$.
The SCF cycle is summarized as $\sigma_q^p \mapsto \{\epsilon_p, \phi_p(\mathbf{x})\} \mapsto T_{ij}^{ab}
 \mapsto \sigma_q^p$.

%P2.10 - MP2-like self-energy
The inner SCF cycle can be avoided by using a first-order
 approximation for $T_{ij}^{ab}$ instead of solving Eq. (\ref{diag_Riccati}),
\begin{equation} \label{T0}
 T_{ij}^{ab0} = \frac{-V_{ij}^{ab}}{\epsilon_a - \epsilon_i + \epsilon_b - \epsilon_j}.
\end{equation}
$T_{ij}^{ab} \Rightarrow T_{ij}^{ab0}$ in Eq. (\ref{sigma}) defines $\sigma_q^p \Rightarrow \sigma_q^{p0}$,
 which is similar to a canonical transformation MP2 (CT-MP2) theory \cite{CT-MP2}.
Other theories use $\sigma_q^{p0}$ to calculate $E$ but
 calculate $\{\epsilon_p, \phi_p(\mathbf{x})\}$ using an inconsistent $\Sigma_q^p$.
Brueckner CC2 (BCC2) theory \cite{BCC2} uses
\begin{equation}
 \Sigma_q^p \Rightarrow \delta_a^p \delta_q^i \sigma_i^{a0} + \delta_i^p \delta_q^a \sigma_i^{a0*},
\end{equation}
 and conventional MP2 theory uses $\Sigma_q^p \Rightarrow 0$.

%P2.11 - Generalized Koopmans' theorem
A generalized Koopmans' theorem \cite{general_Koopmans} applies when
 $E$ and $\{\epsilon_p, \phi_p(\mathbf{x})\}$ are calculated from the same $T_{ij}^{ab}$ and $\Sigma_q^p$,
\begin{align}\label{general_Koopmans}
 \epsilon_a  &= \langle \Phi | \hat{c}_a \exp(-\hat{T}) \hat{H} \exp(\hat{T}) \hat{c}_a^\dag | \Phi \rangle - E, \notag \\
  \epsilon_i &= E - \langle \Phi | \hat{c}_i^\dag \exp(-\hat{T}) \hat{H} \exp(\hat{T}) \hat{c}_i | \Phi \rangle ,
\end{align}
 which generalizes Eq. (\ref{Koopmans}).
We assume $b_q^p$ to be real for $p=q$.
%Assuming $b_q^p$ is real for $p=q$,
%It relates orbital energies to total energy differences as in Eq. (\ref{Koopmans}),
This partially corrects for the absence of electron correlation in Koopmans' theorem,
 but it still has an ``orbital relaxation'' error.
Because solutions to the BRPA equations depend on the choice of orbital occupations,
 $\epsilon_p$ is not exactly the difference between a pair of converged BRPA total energies.
In contrast, many-body Green's function theory is formally exact \cite{GW}.

%S3
\section{Fast algorithm components\label{fast_components}}
\vspace{-7pt}

%P3.1 - None of the components are new
None of the three components of the fast BRPA algorithm are fundamentally new.
However, some details of their use are nonstandard.
We review these details and how they compare with modern standards in
 the Gaussian-orbital and planewave-pseudopotential electronic structure methodologies.

%S3.1
\vspace{-2pt}
\subsection{Primary and auxiliary local basis sets\label{basis_sets}}
\vspace{-7pt}

%P3.2 - Simple choice of primary and auxiliary basis
In this paper, we use a grid in $\mathbf{x}$ with $\alpha n$ points as
 both the primary and auxiliary basis, labeled by $\{x,y,z,w\}$.
$\alpha$ is a basis set efficiency factor.
The relation between primary and orbital fermion operators is 
 $\hat{a}_x = \phi_{px} \hat{c}_p$ with orthonormal structure, $\phi_{px} \phi_{py}^* = \delta_{xy}$.
Single-electron operators in the primary basis are $X_{xy} = \phi_{px} X_q^p \phi_{qy}^*$.
This notation transparently interchanges with basis-free notation (e.g. $\phi_p(\mathbf{x}) \Leftrightarrow \phi_{px}$).
In terms of $\hat{a}_x$,
\begin{equation} \label{aux_H}
 \hat{H} =  E_0 + h_{xy} \hat{a}_x^\dag \hat{a}_y + \tfrac{1}{2}  V_{xy} \hat{n}_x \hat{n}_y - \tfrac{1}{2} V_{xx} \hat{n}_x ,
\end{equation}
 with number operators, $\hat{n}_x = \hat{a}_x^\dag \hat{a}_x$, and a ``kernel'', $V_{xy}$, from
\begin{equation} \label{V_THC}
 V_{rs}^{pq} = S_{rx}^{p0} V_{xy} S_{sy}^{q0}, \ \ \ S_{qx}^{p0} = \phi_{px}^* \phi_{qx}.
\end{equation}
This is similar to a tensor hypercontraction (THC) form \cite{THC} for $V_{rs}^{pq}$.
However, here we use it as a theoretical basis to simplify accounting and not necessarily as a computational basis.

%P3.3 - Requirements of a primary and auxiliary basis
A computational basis must enable decomposition of $V_{rs}^{pq}$ similar to Eq. (\ref{V_THC})
 to be suitable for a fast BRPA calculation.
Its kernel, $V_{xy}$, must be amenable to fast summation methods discussed in Sec. \ref{fastVsum}.
A general form, $S_{qz}^{p0} = \phi_{px}^* S_{xyz}^0 \phi_{qy}$,
 is acceptable if the ``vertex'', $S_{xyz}^0$, is sparse.
In $S_{xyz}^0$, $\{x,y\}$ are primary basis indices, and $z$ is an auxiliary basis index.
These primary and auxiliary bases need not be orthogonal or equal,
 but we assume both to simplify method development.

%P3.4 - Summarize primary/auxiliary basis sets in Gaussian-orbital methodology
In Gaussian-orbital electronic structure, both primary and auxiliary basis functions are
 atom-centered polynomials times Gaussians.
The resolution-of-identity (RI) form of $V_{rs}^{pq}$ is \cite{auxiliary}
\begin{subequations}\begin{align}
 S_{xyz}^0 &= \int d\mathbf{x} d\mathbf{y} f_x^*(\mathbf{x}) f_y(\mathbf{x}) V(\mathbf{x},\mathbf{y}) g_z(\mathbf{y}), \label{RI_vertex} \\
 V_{zw}^{-1} &= \int d\mathbf{x} d\mathbf{y} g_z(\mathbf{x}) V(\mathbf{x},\mathbf{y}) g_w^*(\mathbf{y}), 
\end{align}\end{subequations}
 for primary, $f_x(\mathbf{x})$, and auxiliary, $g_z(\mathbf{x})$, basis functions and a matrix inverse, `$^{-1}$'.
The RI vertex is not sparse, which is also the case in
 Cholesky \cite{Cholesky} and pseudospectral \cite{pseudospectral} decompositions.
The THC vertex is sparse but requires $\mathcal{O}(\alpha^4 n^4)$ operations to be calculated at present \cite{THC}.
In a standard RI-RPA calculation \cite{oldfastRPA},
 $\alpha = 14.5$ for the cc-pVTZ basis and $\alpha = 40.5$ for its auxiliary basis \cite{cc_aux}
 of a frozen-core C atom ($n=4$).

%P3.5 - Summarize primary/auxiliary basis sets in planewave-pseudopotential methodology
In planewave-pseudopotential electronic structure, all the basis functions
 are planewaves, but pseudopotentials augment the primary basis inside atomic spheres \cite{PAW}.
No augmentation of the auxiliary basis is a source of errors when core-valence polarization is important \cite{VASP_RPA}.
Planewaves have a sparse vertex because the fast Fourier transform (FFT) enables efficient grid operations.
Also, the Coulomb kernel, $V_{xy}$, is diagonal in the planewave basis.
In a standard planewave RPA calculation \cite{VASP_RPA},
 $\alpha = 57.3$ for the primary basis and $\alpha = 28.3$ for the auxiliary basis of a frozen-core C atom in diamond at equilibrium.

%S3.2
\subsection{Low-rank energy denominator approximation}
\vspace{-7pt}

%P3.6 - Contour integration form (w/ second splitting)
In this paper, we build low-rank approximations to energy denominators
 using numerical quadratures of three integrals.
The first integral is over the imaginary frequency axis,
\begin{align} \label{quadrature1}
 \frac{1}{\omega_{ai} + \omega_{bj}} &= \int_{-i\infty}^{i\infty} 
 \frac{d\Omega}{2 \pi i} \frac{1}{(\omega_{ai} - \Omega)(\omega_{bj} + \Omega)} \notag \\
 &\approx \frac{\Omega_e}{(\omega_{ai} - \omega_e)(\omega_{bj} + \omega_e)} ,
\end{align}
 with $\omega_{ai} = \epsilon_a - \epsilon_i > 0$ and a quadrature of $\beta_1$ points,
 $\omega_e$, and weights, $\Omega_e$, indexed by $\{e,f,g\}$.
We use a permuted index, $\overline{e}$, to write a required quadrature symmetry:
 $\omega_{\overline{e}} = \omega_e^* = - \omega_e$ and $\Omega_{\overline{e}} = \Omega_e^* = \Omega_e$.
The remaining two integrals are
\begin{align} \label{quadrature2}
 \frac{\delta_a^p \delta_i^q}{\omega_{ai} + \omega_e} &= \oint_{\Gamma_{\mathrm{v}}} \frac{d\Omega}{2\pi i} \oint_{\Gamma_{\mathrm{o}}} 
 \frac{d\Omega'}{2\pi i} \frac{1/(\Omega - \Omega' + \omega_e)}{(\epsilon_p - \Omega)(\epsilon_q - \Omega')} \notag \\
  &\approx \frac{\Omega_{\underline{a}\underline{i}}^e}{(\epsilon_p - \omega_{\underline{a}})(\epsilon_q - \omega_{\underline{i}})},
\end{align}
 with closed counterclockwise contours, $\Gamma_\mathrm{v}$ separating $\epsilon_a$ from $\epsilon_i - \omega_e$, and
 $\Gamma_\mathrm{o}$ separating $\epsilon_i$ from $\epsilon_a+ \omega_e$.
Their quadratures are $\beta_2$ points, $\omega_{\underline{a}}$, indexed by $\{\underline{a},\underline{b}\}$
 and $\beta_3$ points, $\omega_{\underline{i}}$, indexed by $\{\underline{i},\underline{j}\}$
 with different weights, $\Omega_{\underline{a}\underline{i}}^e$, for each $\omega_e$.
We combine $\underline{a}$ and $\underline{i}$ into one index, $\underline{p}$, for notational convenience.

%P3.7 - Analytic contour quadrature: appendix summary & results from others
With a 4-parameter model of the orbital energy spectrum,
 $\epsilon_a \in [ \epsilon_{\mathrm{v}}^{\mathrm{min}} , \epsilon_{\mathrm{v}}^{\mathrm{max}} ]$ and
 $\epsilon_i \in [ \epsilon_{\mathrm{o}}^{\mathrm{min}} , \epsilon_{\mathrm{o}}^{\mathrm{max}} ]$,
 and a target error, $\varepsilon_{\mathrm{Q}}$,
 we analytically construct numerical quadratures in Appendix \ref{quadrature_section}.
Comparable to other results \cite{Lin_quadrature}, the quadrature sizes are
\begin{align}
 \beta_1 &\approx \frac{2}{\pi^2} \ln \left( \frac{\epsilon_{\mathrm{v}}^{\mathrm{max}} - \epsilon_{\mathrm{o}}^{\mathrm{min}}}
 {\epsilon_{\mathrm{v}}^{\mathrm{min}} - \epsilon_{\mathrm{o}}^{\mathrm{max}}} \right) \ln \varepsilon_{\mathrm{Q}}^{-1} , \notag \\
 \beta_2 &\approx \frac{4}{\pi^2} \ln \left( \frac{\epsilon_{\mathrm{v}}^{\mathrm{max}} - \epsilon_{\mathrm{o}}^{\mathrm{max}}}
 {\epsilon_{\mathrm{v}}^{\mathrm{min}} - \epsilon_{\mathrm{o}}^{\mathrm{max}}} \right) \ln \varepsilon_{\mathrm{Q}}^{-1} , \notag \\
 \beta_3 &\approx \frac{4}{\pi^2} \ln \left( \frac{\epsilon_{\mathrm{v}}^{\mathrm{min}} - \epsilon_{\mathrm{o}}^{\mathrm{min}}}
 {\epsilon_{\mathrm{v}}^{\mathrm{min}} - \epsilon_{\mathrm{o}}^{\mathrm{max}}} \right) \ln \varepsilon_{\mathrm{Q}}^{-1} ,
\end{align}
 without an explicit $n$-dependence.
When $\epsilon_{\mathrm{v}}^{\mathrm{min}} - \epsilon_{\mathrm{o}}^{\mathrm{max}}$
 is zero or very small, a low-energy cutoff must be introduced.

%P3.8 - Closure relations
In conjunction with Eq. (\ref{quadrature1}), the fast RPA algorithm
 also requires an approximate product closure relation,
\begin{equation}\label{closure}
  \frac{1}{(\omega_{ai} + \omega_e)(\omega_{ai} + \omega_f)} \approx \frac{\Delta_g^{ef}}{\omega_{ai} + \omega_g}.
\end{equation}
For $e \neq f$, this equation is closed by partial fractions,
\begin{equation}
  \Delta_g^{ef} = - \delta_f^e \nabla_g^f - (\delta_g^e - \delta_g^f) / (\omega_e - \omega_f + \delta_f^e) .
\end{equation}
For $e = f$, this requires coefficients, $\nabla_f^e$, of a finite difference approximation
 to $-\tfrac{d}{d\omega} (\omega_{ai} + \omega)^{-1}$ for $\omega \in \{\omega_e \}$.
It is a standard linear approximation problem that we solve by minimizing a
 root-mean-square (RMS) error metric,
\begin{equation}
 \varepsilon_{\mathrm{FD}} = \min_{\nabla_f^e} \sqrt{ \frac{1}{(\alpha-1) \beta_1 n^2} \sum_{a,i,e} \left| 1 + \nabla_f^e \frac{(\omega_{ai} + \omega_e)^2}{\omega_{ai} + \omega_f} \right|^2} .
\end{equation}
In practice, we find that $\varepsilon_{\mathrm{FD}}$ is proportional to $\varepsilon_{\mathrm{Q}}$.

%P3.9 - Laplace transform
Numerical quadratures of the Laplace transform \cite{Laplace}
 are an alternative low-rank energy denominator approximation,
\begin{align} \label{Laplace_quadrature}
 \frac{1}{\epsilon_a - \epsilon_i + \epsilon_b - \epsilon_j} &=
  \int_0^\infty ds e^{-s \epsilon_a} e^{s \epsilon_i} e^{-s \epsilon_b} e^{s \epsilon_j} \notag \\
  &\approx w_e e^{-s_e \epsilon_a} e^{s_e \epsilon_i} e^{-s_e \epsilon_b} e^{s_e \epsilon_j} .
\end{align}
Numerical methods for calculating quadratures are known \cite{Laplace_quadrature}.
The main advantage of Eq. (\ref{Laplace_quadrature}) over Eqs. (\ref{quadrature1}) and (\ref{quadrature2})
 is the single quadrature summation rather than nested summations.
However, Eqs. (\ref{quadrature1}) and (\ref{quadrature2}) have better numerical behavior
 in the case of complex orbital energies and lead to equations with connections to many-body Green's function theory \cite{GW}.

%P3.10 - Kernel interpolation
Kernel interpolation (i.e. skeleton decomposition \cite{kernel_interp}) 
 is yet another low-rank approximation of energy denominators,
\begin{align}
 \frac{1}{\omega_{ai} + \omega_{bj}} &\approx \frac{1}{\omega_{ai} + \omega_e} [\mathbf{K}^{-1}]_{ef} \frac{1}{\omega_{bj} + \omega_f}, \notag \\
  [\mathbf{K}]_{ef} &= \frac{1}{\omega_e + \omega_f} .
\end{align}
It is not based on reduction of an integral to quadrature, thus it does not have a simple exact limiting expression.
However, it is exact if $\omega_{ai}$ or $\omega_{bj}$ is in $\{\omega_e\}$ and may be efficient 
 for treating isolated spectral features and large interior energy gaps.
 
%S3.3
\vspace{-2pt}
\subsection{Fast summation of the interaction kernel\label{fastVsum}}
\vspace{-7pt}

%P3.11 - Basic accounting rules
In this paper, fast $V_{xy}$ summation methods are accounted for by assigning
 $\mathcal{O}(\alpha \gamma n)$ operations to $\rho_x \mapsto V_{xy} \rho_y$.
$\gamma$ is an efficiency factor with either weak or no $n$-dependence for fast methods.
Further details are not important for cost analysis.

%P3.12 - Use of fast summation methods in electronic structure
Fast $V_{xy}$ summation methods are ubiquitous in electronic structure.
FFTs are used in planewave-pseudopotential codes to solve the Poisson equation
 and apply the Hamiltonian to an orbital.
The fast multiple method (FMM) is used in Gaussian-orbital codes
 to efficiently calculate the matrix elements of the Hartree potential \cite{CFMM}.
Except for planewave calculations of the Fock exchange \cite{PW_exchange},
 these fast methods are not a bottleneck and their $\gamma$ values do not
 contribute to the leading-order cost.

%P3.13 - Survey of fast summation methods
The leading-order cost of the fast BRPA algorithm has a dependence on $\gamma$.
This is commonly the case for classical $n$-body simulations of electrostatics in molecular dynamics
 and gravity in astrophysics.
As a result, the method development in those fields has been driven to develop new fast summation techniques.
Such modern innovations include the accelerated cartesian expansion \cite{ACE} (ACE),
 multilevel summation method \cite{MSM} (MSM), and 
 hierarchical matrix decompositions \cite{Hmat}.
However, electronic structure applications have special requirements for summation methods
 such as the ability to treat point nuclei, smooth valence electron charge,
 and the multiple length scales of core electron charge in a unified framework.
Applications such as BRPA might motivate this type of development.

%P3.14 - Other structure in the Coulomb interation
Fast summation methods exist for other $V_{xy}$ besides the Coulomb kernel.
The fast BRPA algorithm applies to any $V_{xy}$ for which $\gamma$ is small for any reason.
Such generality might be appealing, but further improvements to accuracy or efficiency
 could result from better use of structure within the Coulomb kernel
 and elementary operations beyond just $\rho_x \mapsto V_{xy} \rho_y$.
An example is the low off-diagonal rank of the Coulomb kernel shared by a general class of structured matrices \cite{Hmat}.

%S4
\section{RPA tensor structure\label{RPAstructure}}
\vspace{-7pt}

%P4.1 - Rich structure of equivalences in RPA
There are three distinct approaches to calculating the RPA correlation energy.
The first is calculating $T_{ij}^{ab}$ by solving
 the RPA Riccati equation in Eq. (\ref{BRPA_double})
 and evaluating
\begin{equation}
 E_{\mathrm{c}}^{\mathrm{RPA}} = \tfrac{1}{2} V_{ab}^{ij} T_{ij}^{ab} .
\end{equation}
The RPA Riccati equation is equivalent to the RPA symplectic eigenvalue problem \cite{RPAfromCC}
 that determines electron-hole excitation energies.
 $E_{\mathrm{c}}^{\mathrm{RPA}}$ is also half the sum of the difference between
 these excitation energies and the corresponding energies in the
 Tamm-Dancoff approximation \cite{RPAfromACFD}.
These energies are the poles of frequency-dependent polarization functions
 and this sum is extracted from them by the residue theorem when used on the
 adiabatic-connection fluctuation-dissipation (ACFD) formula for $E_{\mathrm{c}}^{\mathrm{RPA}}$.
This formula can be rearranged as
\begin{equation} \label{ACFD_E}
 E_{\mathrm{c}}^{\mathrm{RPA}} = \tfrac{1}{4} V_{pq}^{rs} C_{rs}^{pq} ,
\end{equation}
 where $C_{rs}^{pq}$ is the correlated part of the RPA two-body density matrix averaged over interaction strength \cite{RPA_equiv}.
We use the well-studied structure of $C_{rs}^{pq}$ as a reference point
 for the discussion of structure in $T_{ij}^{ab}$, which has yet to be elucidated.

\vspace{-2pt}
\subsection{Adiabatic-connection fluctuation-dissipation RPA}
\vspace{-7pt}

%P4.2 - Direct ACFD formula for C & convenient arrangements
In an auxiliary basis, the ACFD formula for $E_{\mathrm{c}}^{\mathrm{RPA}}$ is \cite{RPAfromACFD}
\begin{equation} \label{ACFD_RPA}
 E_{\mathrm{c}}^{\mathrm{RPA}} = - \int_0^1 d \lambda \int_{-i \infty}^{i \infty} \frac{d \Omega}{4 \pi i} V_{xy} [ P_{xy}^{\lambda}(\Omega) - P_{xy}^0(\Omega) ] ,
\end{equation}
 where $P_{xy}^{\lambda}(\omega)$ is the RPA polarization function for interaction strength $\lambda$.
 $P_{xy}^{\lambda}(\omega)$ is defined in relation to its $\lambda=0$ value as
\begin{align}
 P_{xy}^0(\omega) &= - \frac{S_{ix}^{a0}S_{ay}^{i0}}{\omega_{ai} + \omega} - \frac{S_{ax}^{i0}S_{iy}^{a0}}{\omega_{ai} - \omega} , \notag \\
 P_{xy}^{\lambda}(\omega) &= P_{xy}^0(\omega) + \lambda P_{xz}^{\lambda}(\omega) V_{zw} P_{wy}^0(\omega) . \label{P_def}
 \end{align}
The screened Coulomb interaction is related to $P_{xy}^{\lambda}(\omega)$ as
\begin{subequations}\begin{align}
 W_{xy}^{\lambda}(\omega) &=  \lambda V_{xy} + \lambda^2 V_{xz} P_{zw}^{\lambda}(\omega) V_{wy} , \label{W_def} \\
 P_{xy}^{\lambda}(\omega)  &= P_{xy}^0(\omega) + P_{xz}^0(\omega) W_{zw}^{\lambda}(\omega) P_{wy}^0(\omega). \label{W_def2}
\end{align}\end{subequations}
Using Eq. (\ref{W_def2}), we encapsulate $\lambda$ in Eq. (\ref{ACFD_RPA}) with
\begin{align}
 \overline{W}_{xy}(\omega) &= 2 \int_0^1 d \lambda W_{xy}^{\lambda}(\omega) , \ \ \ [\mathbf{E}(\omega)]_{xy} = P_{xz}^0(\omega) V_{zy} , \notag \\
 \overline{W}_{xy}(\omega) &= - 2 V_{xz} [\mathbf{E}(\omega)^{-1} + \mathbf{E}(\omega)^{-2} \ln(\mathbf{I} - \mathbf{E}(\omega))]_{zy} ,
\end{align}
 and extract $C_{rs}^{pq}$ by splitting $P_{xy}^0(\omega)$ over its internal indices,
\begin{align} \label{C_def}
 C_{rs}^{pq} &= - \delta_a^p \delta_b^q \delta_r^i \delta_s^j \int_{-i \infty}^{i \infty} \frac{d\Omega}{2 \pi i}
  \frac{S_{ix}^{a0} \overline{W}_{xy}(\Omega) S_{jy}^{b0} }{(\omega_{ai} - \Omega)(\omega_{bj} + \Omega)} \notag \\
  & \ \ \ - \delta_i^p \delta_j^q \delta_r^a \delta_s^b \int_{-i \infty}^{i \infty} \frac{d\Omega}{2 \pi i} 
 \frac{S_{ax}^{i0} \overline{W}_{xy}(\Omega) S_{by}^{j0} }{(\omega_{ai} + \Omega)(\omega_{bj} - \Omega)} .
\end{align}
Other forms \cite{oldfastRPA,SOSEX_implementation,RPA_equiv} of $C_{rs}^{pq}$
 combine the `$_{ij}^{ab}$' and `$_{ab}^{ij}$' sectors and
 use different but equivalent energy denominators.

\subsection{Structure of $T_{ij}^{ab}$ in BRPA\label{T_struct}}
\vspace{-7pt}

%P4.3 - Define the S-form of T
From observing Eqs. (\ref{diag_Riccati}), (\ref{quadrature1}), and (\ref{C_def}),
 it is reasonable to expect similar tensor structure in $C_{rs}^{pq}$ and $T_{ij}^{ab}$.
To that end, we expand $V_{rs}^{pq}$ in Eq. (\ref{diag_Riccati}) using Eq. (\ref{V_THC}) and regroup terms,
\begin{equation}\label{T_factor}
 T_{ij}^{ab} = \frac{- S_{ix}^a V_{xy} S_{jy}^b}{\omega_{ai} + \omega_{bj}}, \ \ \ 
  S_{ix}^a = S_{ix}^{a0} + T_{ij}^{ab} S_{bx}^{j0} .
\end{equation}
This can be rewritten to define $S_{ix}^a$ without reference to $T_{ij}^{ab}$,
\begin{subequations} \label{almost_Dyson} \begin{align}
 S_{ix}^a &= S_{ix}^{a0} + S_{iy}^a V_{yz} B_{zx}(\omega_{ai}) , \label{S_def} \\
 B_{xy}(\omega) &= - \frac{S_{ix}^a S_{ay}^{i0}}{\omega_{ai} + \omega} .
\end{align} \end{subequations}
The iteration of Eq. (\ref{S_def}) starting from $S_{ix}^a = S_{ix}^{a0}$ elucidates a factorization ansatz that further simplifies $S_{ix}^a$,
\begin{equation} \label{A_ansatz}
 S_{ix}^a = S_{ix}^{a0} + S_{iy}^{a0} V_{yz} A_{zx}(\omega_{ai}) .
\end{equation}
It enables a reduction of Eq. (\ref{almost_Dyson}) to Dyson-like equations,
\begin{subequations}\label{new_Dyson}\begin{align}
 A_{xy}^0(\omega) &= - \frac{S_{ix}^{a0} S_{ay}^{i0}}{\omega_{ai} + \omega} , \label{A0_def} \\
 A_{xy}(\omega) &= B_{xy}(\omega) + A_{xz}(\omega) V_{zw} B_{wy}(\omega) , \label{A_def} \\
 B_{xy}(\omega) &= A_{xy}^0(\omega) - \oint_\Gamma \frac{d\Omega}{2\pi i} A_{zx}(-\Omega)V_{zw} A_{wy}^0[\Omega,\omega] . \label{B_def}
\end{align}\end{subequations}
This uses a divided difference, $f[x,y] = [ f(x) - f(y) ] / (x-y)$, and
 an analytic reconstruction of $A_{xy}(\omega_{ai})$ with the form
\begin{equation} \label{reconstruct}
  A_{xy}(\omega_{ai}) = \oint_\Gamma \frac{d\Omega}{2\pi i} \frac{A_{xy}(-\Omega)}{\omega_{ai} + \Omega}
\end{equation}
 for a closed counterclockwise contour $\Gamma$ separating $-\omega_{ai}$ from the poles of $A_{xy}(-\omega)$,
 which are disjoint if $A_{xy}(\omega_{ai})$ is finite.

%P4.4 - Emphasize inequivalence of C & T
We maximize the superficial similarity of $C_{rs}^{pq}$ and $T_{ij}^{ab}$ by
 again regrouping terms in $T_{ij}^{ab}$ and applying Eq. (\ref{quadrature1}),
\begin{subequations} \begin{align}
 T_{ij}^{ab} &= - \int_{-i \infty}^{i \infty} \frac{d\Omega}{2 \pi i}
  \frac{S_{ix}^{a0} U_{xy}(\omega_{ai},\omega_{bj}) S_{jy}^{b0} }{(\omega_{ai} - \Omega)(\omega_{bj} + \Omega)} , \\
  U_{xy}(\omega,\omega') &= V_{xy} + V_{xz} Q_{zw}(\omega,\omega') V_{wy}, \\
  Q_{xy}(\omega,\omega') &= A_{xy}(\omega) + A_{yx}(\omega') + A_{xz}(\omega) V_{zw} A_{yw}(\omega').
\end{align} \end{subequations}
While both $C_{rs}^{pq}$ and $T_{ij}^{ab}$ produce the same value of $E_{\mathrm{c}}^{\mathrm{RPA}}$,
 their corresponding SOSEX-like values, $E_{\mathrm{c}}^{\mathrm{SOSEX}} = -\tfrac{1}{2} V^{ij}_{ba} T_{ij}^{ab}$
 and $E_{\mathrm{c}}^{\mathrm{AC-SOSEX}} = -\tfrac{1}{4} V_{qp}^{rs} C_{rs}^{pq}$, are not equal
 and begin to differ at third order in perturbation theory \cite{RPA_equiv}.
However, $E_{\mathrm{c}}^{\mathrm{AC-SOSEX}}$ and $E_{\mathrm{c}}^{\mathrm{SOSEX}}$
 are numerically similar for small molecules \cite{SOSEX_implementation}.

%P4.5 - Use of T structure as a basis for algorithm design
In a BRPA calculation based on Sec. \ref{SCF_BRPA}, $T_{ij}^{ab}$ will appear in its own calculation and also in the calculation of $\sigma_q^p$.
There are $\mathcal{O}(\alpha^2 n^4)$ variables in $T_{ij}^{ab}$, which is the memory bottleneck.
Direct calculation of $S_{ix}^a$ in Eq. (\ref{almost_Dyson}) and of $\sigma_q^p$ from $S_{ix}^a$
 will reduce the number of variables to $\mathcal{O}(\alpha^2 n^3)$ if the full storage of $B_{xy}(\omega_{ai})$ is avoided.
$A_{xy}(\omega_{ai})$ is not directly useful because it contains $\mathcal{O}(\alpha^3 n^4)$ variables.
Nevertheless, the interpolation of $A_{xy}(\omega_{ai})$
 from an effective quadrature of Eq. (\ref{reconstruct})
 reduces it to $\mathcal{O}(\alpha^2 \beta_1 n^2)$ variables and enables fast algorithms.

%S5
\section{Fast algorithm design\label{fast_design}}
\vspace{-7pt}

%P5.1 - Summarize section
Here we design fast and conventional algorithms for MP2 and RPA+SOSEX calculations.
In each of these four designs, we begin by identifying the tensor equations to be evaluated.
Many of these equations are sufficiently complicated that an algorithm for efficient evaluation
 is not immediately obvious.
We use simple pseudocode to decompose these equations into intermediate variables and elementary operations.
Algorithm costs are then easy to account for in the pseudocode.
Without the simple primary and auxiliary basis sets in Sec. \ref{basis_sets}, this
 design process becomes significantly more difficult.
Even so, the results here demonstrate what is possible with a practical basis set
 if sufficient effort is given to algorithm design.

%P5.2 - Define pseudocode syntax
The pseudocode is grouped into functions.
Memory usage is delineated at the beginning of the function declaration with
 inputs then outputs separated by a semicolon in the function argument
 and a workspace statement that lists all temporary variables.
The body of each function contains `for' loops and elementary `$A := B$' operations
 that denote the calculation of expression $B$ and its storage in variable $A$.
Each bottleneck is commented with its leading-order operation count in the cost polynomial
 of $\{\alpha,\beta_1,\beta_2,\beta_3,\gamma,n\}$.
We count as one operation each addition, multiplication, and division.
Real and complex are not distinguished in operation and variable counts.

%P5.3 - Define design process: constraints & goals
The basic design strategies for minimizing cost are reuse of calculations
 and avoidance of concurrent data storage.
All computational costs that are leading order in $n$
 are arranged as either matrix-matrix multiplications with the minimum matrix dimension maximized
 or as fast $V_{xy}$ summations.
This choice hides the cost of data movement in efficient implementations of linear algebra and $V_{xy}$ summation.
These strategies suffice for the design of serial algorithms, but parallel algorithms will
 need to explicitly account for communication costs.

%P5.4 - Common primary-basis SCF structure
A common element of the four algorithms is calculation of $\sigma_q^{p0}$ or $\sigma_q^p$ directly in the primary basis.
The SCF equations from Sec. \ref{SCF_BRPA} in the primary basis are analogous to Eq. (\ref{SCF_space}), 
\begin{subequations} \label{main} \begin{align}
  E &= E_0 + \tfrac{1}{2} \rho_{xy} ( h_{yx} + f_{yx} ) , \\
 f_{xy} &= h_{xy} + v_x \delta_{xy} - \rho_{xy} V_{xy} + \Sigma_{xy} , \\
f_{xy} \phi_{py} &= \epsilon_p \phi_{px} , \\
 \rho_{xy} &= \phi_{ix} \phi_{iy}^* , \\
 v_x &= V_{xy} \rho_{yy} , \\
 \Sigma_{xy} &=  \tfrac{1}{2} ( \sigma_{xy} + \sigma_{yx}^* + \sigma_{xz} \rho_{zy} + \rho_{xz} \sigma_{yz}^* \notag \\
  & \ \ \ \ \ \ \ \ - \rho_{xz} \sigma_{zw} \rho_{wy} - \rho_{xz} \sigma_{wz}^* \rho_{wy} ) , \\
  \sigma_{xy} &= \phi_{ax} \phi_{iy}^* \widetilde{T}_{ij}^{ab} ( f_b^{j0} + V_{xz} S_{bz}^{j0} - S_{bz}^{j0} V_{zy} ) , \label{sigma_xy}
\end{align} \end{subequations}
 with $E$ and $\Sigma_q^p$ also calculated in the primary basis.

%S5.1
\vspace{-2pt}
\subsection{Conventional MP2 algorithm}
\vspace{-7pt}

%P5.5 - Comparison to RI-MP2
Algorithm \ref{MP2c} calculates $\sigma_{xy}^0$ using Eqs. (\ref{T0}) and (\ref{sigma_xy}) by substituting
 $\sigma_{xy} \Rightarrow \sigma_{xy}^0$ and $T_{ij}^{ab} \Rightarrow T_{ij}^{ab0}$.
It is comparable to an RI-MP2 algorithm \cite{RI_MP2} when
 Eq. (\ref{main}) is used to calculate $E$ from $\sigma_{xy}^0$.
Both require $\mathcal{O}(\alpha^3 n^5)$ operations, but
 the $\mathcal{O}(\alpha^2 n^3)$ memory cost of RI-MP2 reduces to an $\mathcal{O}(\alpha^2 n^2)$
 memory cost by avoiding storage of the dense RI vertex in Eq. (\ref{RI_vertex}).
The intermediate calculation of $\sigma_{xy}^0$ instead of a direct calculation of $E$
 enables the self-consistent MP2 methods in Sec. \ref{SCF_BRPA}.

%A1 - Conventional MP2
\begin{algorithm}[h!]
\caption{Conventional MP2\label{MP2c}}
\begin{algorithmic}[1]
\Function {MP2c}{$f_a^{i0}, V_{xy}, \epsilon_p, \phi_{px}; \sigma_{xy}^0$}
 \State $\mathbf{workspace} : \{ X, X_x, X_i^a, X_{ix}, Y_{ix} \}$
\State $\sigma_{xy}^0 := 0$
\ForAll{$a$}
 \State $X_{ix} := \phi_{ix} \phi_{ax}^*$ 
 \State $X_{ix} := V_{xy} X_{iy}$ 
 \ForAll{$i$}
  \State $Y_{jx} := X_{jx} \phi_{ix} - X_{ix} \phi_{jx}$
  \State $X_j^b := \phi_{bx}^* Y_{jx}$ \Comment{$2 \alpha^3 n^5$}
  \State $X_j^b := X_j^b / (\epsilon_a - \epsilon_i + \epsilon_b - \epsilon_j)$
  \State $X := X_j^b f_b^{j0}$
  \State $Y_{jx} := X_j^b \phi_{bx}$ \Comment{$2 \alpha^3 n^5$}
  \State $X_x := \phi_{jx}^* Y_{jx}$
  \State $X_x := V_{xy} X_y$
  \State $\sigma_{xy}^0 := \sigma_{xy}^0 + \phi_{ax} \phi_{iy}^* ( X + X_x - X_y )$
 \EndFor
\EndFor
\EndFunction
\end{algorithmic}
\end{algorithm}

%S5.2
\vspace{-2 pt}
\subsection{Conventional BRPA algorithm\label{BRPAc_section}}
\vspace{-7pt}

%P5.6 - Definition of residual in the inner loop
Because of the $T_{ij}^{ab}$ structure found in Sec. \ref{T_struct},
 the SCF inner loop proposed in Sec. \ref{SCF_BRPA} only needs to determine $S_{ix}^a$ rather than $T_{ij}^{ab}$.
We rearrange Eq. (\ref{almost_Dyson}) into a residual tensor,
\begin{equation}
  R_{ix}^a = S_{ix}^a - S_{ix}^{a0} + S_{iy}^a \frac{V_{yz} S_{jz}^b S_{bx}^{j0}}{\omega_{ai} + \omega_{bj}},
\end{equation}
 and recast the SCF inner loop as solving $R_{ix}^a = 0$.
Algorithm \ref{inBRPAc} calculates $R_{ix}^a$ with $\mathcal{O}(\alpha^3 n^5)$ operations and $\mathcal{O}(\alpha^2 n^3)$ memory.
As in the MP2 case, the operation count is the same as other RPA+SOSEX algorithms \cite{VASP_SOSEX},
 but avoiding the storage of $T_{ij}^{ab}$ reduces the $\mathcal{O}(\alpha^2 n^4)$ memory cost.

%P5.7 - Outer loop & balance of cost determining SCF cycle
Algorithm \ref{outBRPAc} calculates $\sigma_{xy}$ using Eqs. (\ref{T_factor}) and (\ref{sigma_xy}).
It is comparable in cost with the $R_{ix}^a$ inner loop calculations, which
 violates the assumption in Sec. \ref{SCF_BRPA} of an inexpensive inner loop.
As a result, it is more efficient to calculate $R_{ix}^a$ and $\sigma_{xy}$ concurrently
 and solve the orbital self-consistency and $R_{ix}^a = 0$ problems simultaneously.
There will be one iterative cycle as opposed to inner iterations nested within outer iterations.
This is an example of modifying the overall design of an algorithm based on a change in the relative cost of its components.

%A2 - Conventional BRPA inner loop
\begin{algorithm}[h!]
\caption{Conventional BRPA inner loop\label{inBRPAc}}
\begin{algorithmic}[1]
\Function{inBRPAc}{$V_{xy}, \epsilon_p, \phi_{px}, S_{ix}^a ; R_{ix}^a$}
 \State $\mathbf{workspace} : \{ X_i^a, X_{ix} , Y_{ix} \}$
 \State $R_{ix}^a := S_{ix}^a - \phi_{ax}^* \phi_{ix}$
 \ForAll{$a$}
  \State $X_{ix} := V_{xy} S_{iy}^a$
 \ForAll{$i$}
  \State $X_j^b := S_{ix}^b X_{jx}$  \Comment{$2 \alpha^3 n^5$}
  \State $X_j^b := X_j^b / (\epsilon_{a} - \epsilon_{i} + \epsilon_{b} - \epsilon_{j})$
  \State $Y_{jx} := X_j^b \phi_{bx}$  \Comment{$2 \alpha^3 n^5$}
  \State $R_{jx}^a := R_{jx}^a + Y_{jx} \phi_{ix}^*$
   \EndFor
 \EndFor
\EndFunction
\end{algorithmic}
\end{algorithm}

%A3 - Conventional BRPA outer loop
\begin{algorithm}[h!]
\caption{Conventional BRPA outer loop\label{outBRPAc}}
\begin{algorithmic}[1]
\Function{outBRPAc}{$f_a^{i0}, V_{xy}, \epsilon_p, \phi_{px}, S_{ix}^a ; \sigma_{xy}$}
\State $\mathbf{workspace} : \{X_i^a, Y_i^a, X_{ix},Y_{ix}, X_{ix}^a \}$
\State $X_i^a  := 0$
\State $X_{ix}^a := 0$
\ForAll{$a$}
 \State $X_{ix} := V_{xy} S_{iy}^a$
 \ForAll{$i$}
  \State $Y_j^b := S_{ix}^b X_{jx}$   \Comment{$2 \alpha^3 n^5$}
  \State $Y_j^b := Y_j^b / (\epsilon_a - \epsilon_i + \epsilon_b - \epsilon_j)$
  \State $X_j^a := X_j^a - Y_j^b f_b^{i0}$
  \State $X_i^a := X_i^a +  Y_j^b f_b^{j0}$
  \State $Y_{jx} := Y_j^b \phi_{bx}$  \Comment{$2 \alpha^3 n^5$}
  \State $X_{jx}^a := X_{jx}^a - Y_{jx} \phi_{ix}^*$
  \State $X_{ix}^a := X_{ix}^a + Y_{jx} \phi_{jx}^*$
 \EndFor
\EndFor
 \State $X_{ix}^a := V_{xy} X_{iy}^a$ 
 \State $\sigma_{xy} := \phi_{ax} \phi_{iy}^* ( X_i^a + X_{ix}^a - X_{iy}^a )$
\EndFunction
\end{algorithmic}
\end{algorithm}

%S5.3
\subsection{Fast MP2 algorithm}
\vspace{-7pt}

%P5.8 - Relation to fast BRPA case, splitting of A0
MP2 corresponds to $A_{xy}(\omega) = 0$ in terms of the structure discussed in Sec. \ref{T_struct}.
However, $A_{xy}^0(\omega_{ai})$ is still an element of fast MP2 calculations.
To efficiently calculate $A_{xy}^0(\omega_{ai})$,
 we introduce a mean-field Green's function,
\begin{equation}
 G_{xy}(\omega) = \frac{\phi_{px} \phi_{py}^*}{\omega - \epsilon_p} , \ \ \ G_{xy}^{\underline{p}} = G_{xy}(\omega_{\underline{p}}) .
\end{equation}
With Eqs. (\ref{quadrature1}) and (\ref{quadrature2}), we reduce $A_{xy}^0(\omega_{ai})$ in Eq. (\ref{A0_def}) to
\begin{subequations} \label{A0_simple} \begin{align}
 A_{xy}^0(\omega_{ai}) &\approx \frac{\Omega_e A_{xy}^{e0}}{\omega_{ai} - \omega_e} , \ \ \  A_{xy}^{e0} = A_{xy}^0(\omega_e), \label{A0_split} \\
 A_{xy}^0(\omega_e) &\approx - \Omega_{\underline{a}\underline{i}}^e G_{xy}^{\underline{i}} G_{yx}^{\underline{a}} .
\end{align} \end{subequations}
Given $G_{xy}^{\underline{p}}$, we need $\mathcal{O}(\alpha^2 \beta_1 \beta_2 \beta_3 n^2)$ operations to form $A_{xy}^{e0}$.

%P5.9 - Decomposition of the self-energy
Repeated application of Eqs. (\ref{quadrature1}) and (\ref{quadrature2}) to $\sigma_{xy}^0$ enables
 its decomposition into $G_{xy}^{\underline{p}}$ and rearrangement into
\begin{align}
 \sigma_{xy}^0 &\approx 
 F_{xz}^{\underline{i}0} G_{zy}^{\underline{i}} - G_{xz}^{\underline{a}} F_{zy}^{\underline{a}0} , \notag \\
 F_{xy}^{\underline{i}0} &= \Omega_{\underline{a}\underline{i}}^e G_{xy}^{\underline{a}} W_{yx}^{e0} +
 \Omega_{\underline{a} \underline{i}}^e G_{xz}^{\underline{a}} V_{zy} ( \Xi_{zyx}^{e0} + J_{zy}^{e0}) , \notag \\
 F_{xy}^{\underline{a}0} &= \Omega_{\underline{a}\underline{i}}^e G_{xy}^{\underline{i}} ( W_{xy}^{e0} + v_x^{e0} ) +
 \Omega_{\underline{a} \underline{i}}^e G_{zy}^{\underline{i}} V_{xz} \Xi_{xzy}^{e0} , \notag \\
 \Xi_{xyz}^{e0} &= \Omega_e \Omega_{\underline{a}\underline{i}}^{\overline{e}} G_{xw}^{\underline{i}} G_{wy}^{\underline{a}} V_{wz} , \notag \\
 J_{xy}^{e0}  &=  \Omega_e \Omega_{\underline{a}\underline{i}}^{\overline{e}} G_{xz}^{\underline{i}} f_{zw}^0 G_{wy}^{\underline{a}} , \notag \\
 W_{xy}^{e0} &= \Omega_e V_{xz} A_{zw}^{\overline{e}0} V_{wy} , \notag \\
   v_x^{e0} &= V_{xy} J_{yy}^{e0} . \label{fast_sigma0}
\end{align}
$\sigma_{xy}^0$ is calculated by Algorithm \ref{MP2f} in $\mathcal{O}(\alpha^3 \beta_1 (\beta_2 + \beta_3 + \gamma) n^3)$ operations 
 and $\mathcal{O}(\alpha^2 ( \beta_1 + \beta_2 + \beta_3 ) n^2)$ memory.
This improves on the $\mathcal{O}(\alpha^3 n^4)$ operations used by THC-MP2 \cite{THC_MP2}
 and $\mathcal{O}(\alpha^4 n^4)$ operations used by atomic-orbital Laplace MP2 with integral screening \cite{Haser_MP2}.
Neither method uses fast $V_{xy}$ summation, which enables the extra factor of $n$ speedup
 in the new algorithm.

%This improves on the $\mathcal{O}(\alpha^3 n^4)$ operations used by THC-MP2 \cite{THC_MP2}
% and $\mathcal{O}(\alpha^4 n^4)$ operations used by atomic-orbital Laplace MP2 \cite{Haser_MP2}.
%Neither of these methods use fast $V_{xy}$ summation, which is what enables the additional factor of $n$ speedup
% in the new algorithm.

%A4 - Fast MP2
\begin{algorithm}[h!]
\caption{Fast MP2\label{MP2f}}
\begin{algorithmic}[1]
\Function {MP2f}{$\Omega_e, \Omega_{\underline{a}\underline{i}}^e, f_{xy}^0, V_{xy}, G_{xy}^{\underline{p}}; \sigma_{xy}^0$}
 \State $\mathbf{workspace} : \{ v_x^{e0}, X_x^{e\underline{p}}, J_{xy}^{e0}, W_{xy}^e, X_{xy}^e, F_{xy}^{\underline{p}0} \}$
\State $J_{xy}^{e0} := 0$
\State $W_{xy}^{e0} := 0$
\State $F_{xy}^{\underline{i}0} := G_{xz}^{\underline{i}} f_{zy}^0$
%Calculate J, W, & v
\ForAll{$\underline{i}$}
 \State $X_{xy}^e := \Omega_e \Omega_{\underline{a}\underline{i}}^{\overline{e}} G_{xy}^{\underline{a}}$
  \Comment{$2 \beta_1 \beta_2 \beta_3 \alpha^2 n^2$}
 \State $W_{xy}^{e0} := W_{xy}^{e0} - G_{xy}^{\underline{i}} X_{yx}^e$
 \State $J_{xy}^{e0} := J_{xy}^{e0} + F_{xz}^{\underline{i}0} X_{zy}^e$ \Comment{$2 \beta_1 \beta_3 \alpha^3 n^3$}
\EndFor
\State $W_{xy}^{e0} := V_{xz} W_{zy}^{e0}$
\State $W_{xy}^{e0} := V_{yz} W_{xz}^{e0}$
\State $v_x^{e0} := V_{xy} J_{yy}^{e0}$
%Calculate F
\State $F_{xy}^{\underline{p}0} := 0$
\ForAll{$x$}
 \State $X_y^{e\underline{a}} := \Omega_{\underline{a} \underline{i}}^e G_{xy}^{\underline{i}}$  \Comment{$2 \beta_1 \beta_2 \beta_3 \alpha^2 n^2$}
 \State $X_y^{e\underline{i}}  := \Omega_{\underline{a} \underline{i}}^e G_{yx}^{\underline{a}}$  \Comment{$2 \beta_1 \beta_2 \beta_3 \alpha^2 n^2$}
 \State $F_{xy}^{\underline{a}0} := F_{xy}^{\underline{a}0} + X_y^{e\underline{a}} ( W_{xy}^{e0} + v_x^{e0} )$
 \State $F_{yx}^{\underline{i}0} := F_{yx}^{\underline{i}0} + X_y^{e\underline{i}} W_{xy}^{e0}$
 \State $X_{zy}^e := X_y^{\overline{e}\underline{i}} G_{zy}^{\underline{i}}$ \Comment{$2 \beta_1 \beta_3 \alpha^3 n^3$}
 \State $X_{zy}^e := V_{yw} X_{zw}^e$ \Comment{$\beta_1 \gamma \alpha^3 n^3$}
 \State $X_{zy}^e := \Omega_e X_{zy}^e V_{xz}$
 \State $F_{zy}^{\underline{a}0} := F_{zy}^{\underline{a}0} + X_y^{e\underline{a}} X_{zy}^e$ \Comment{$2 \beta_1 \beta_2 \alpha^3 n^3$}
 \State $X_{zy}^e := X_z^{\overline{e}\underline{a}} G_{zy}^{\underline{a}}$ \Comment{$2 \beta_1 \beta_2 \alpha^3 n^3$}
 \State $X_{zy}^e := V_{zw} X_{wy}^e$ \Comment{$\beta_1 \gamma \alpha^3 n^3$}
 \State $X_{zy}^e := ( \Omega_e X_{zy}^e + J_{xy}^{e0}) V_{xy}$
 \State $F_{zy}^{\underline{i}0} := F_{zy}^{\underline{i}0} + X_z^{e\underline{i}} X_{zy}^e$ \Comment{$2 \beta_1 \beta_3 \alpha^3 n^3$}
\EndFor
 \State $\sigma_{xy}^0 := F_{xz}^{\underline{i}0} G_{zy}^{\underline{i}} - G_{xz}^{\underline{a}} F_{zy}^{\underline{a}0}$
\EndFunction
\end{algorithmic}
\end{algorithm}

%S5.4
\subsection{Fast BRPA algorithm}
\vspace{-7pt}

%P5.10 - Inner loop derivation
As in Eq. (\ref{A0_split}), we use Eq. (\ref{quadrature1}) to approximate $B_{xy}(\omega_{ai})$ with $B_{xy}^e = B_{xy}(\omega_e)$
 and postulate a similar form for $A_{xy}(\omega_{ai})$,
\begin{equation}
 B_{xy}(\omega_{ai}) \approx \frac{\Omega_e B_{xy}^e}{\omega_{ai} - \omega_e} , \ \ \
 A_{xy}(\omega_{ai}) \approx \frac{\Omega_e A_{xy}^e}{\omega_{ai} - \omega_e} ,
\end{equation}
 with $A_{xy}^e \neq A_{xy}(\omega_e)$.
Using Eq. (\ref{closure}), this reduces Eq. (\ref{new_Dyson}) to
\begin{subequations}\begin{align}
 A_{xy}^{e0} &= - \Omega_{\underline{a}\underline{i}}^e G_{xy}^{\underline{i}} G_{yx}^{\underline{a}} , \\
  A_{xy}^e &= B_{xy}^e + A_{xz}^e V_{zw} B_{wy}^e  - \Omega_f \Delta_g^{ef} A_{xz}^f V_{zw} B_{wy}^g , \\
 B_{xy}^e &= A_{xy}^{e0} + \Omega_f \Delta_g^{ef} A_{zx}^{\overline{f}} V_{zw} A_{wy}^{g0} ,
\end{align}\end{subequations}
 for $\omega \in \{ \omega_{ai} \}$.
These equations are equivalent to $R_{xy}^e = 0$ for
\begin{equation}
  R_{xy}^e = A_{xy}^e - B_{xy}^e - A_{xz}^e V_{zw} B_{wy}^e + \Omega_f \Delta_g^{ef} A_{xz}^f V_{zw} B_{wy}^g .
\end{equation}
$R_{xy}^e$ is calculated by Algorithm \ref{inBRPAf} in $\mathcal{O}(\alpha^3 \beta_1 n^3)$ operations and
 $\mathcal{O}(\alpha^2 \beta_1 n^2)$ memory.
Compared to the conventional algorithm in Sec. \ref{BRPAc_section},
 this is more efficient in operations when $\beta_1 \ll n^2$ and in memory when $\beta_1 \ll n$.
The calculation of $\overline{W}_{xy}(\omega)$ at quadrature points in Eq. (\ref{C_def})
 is noniterative and has the same cost, but we lack efficient formulas relating $\overline{W}_{xy}(\omega)$ to $\sigma_{xy}$.

%A5 - Fast BRPA inner loop
\begin{algorithm}[h!]
\caption{Fast BRPA inner loop\label{inBRPAf}}
\begin{algorithmic}[1]
\Function{inBRPAf}{$\omega_e, \Omega_e, \nabla_f^e, V_{xy}, A_{xy}^{e0}, A_{xy}^e ; R_{xy}^e$}
\State $\mathbf{workspace} : \{ X_f^e, X_{xy}^e , Y_{xy}^e \}$
%B calculation (stored in R)
 \State $X_f^e := \Omega_f (1 - \delta_f^e) / (\omega_e - \omega_f + \delta_f^e )$
\State $X_{xy}^e := V_{xz} A_{zy}^{e0}$  \Comment{$ \beta_1 \gamma \alpha^2 n^2$}
 \State $Y_{xy}^e :=  \Omega_e \nabla_f^e X_{xy}^f$ \Comment{$2 \beta_1^2 \alpha^2 n^2$}
 \State $R_{xy}^e := A_{xy}^{e0} - A_{zx}^{\overline{e}} Y_{zy}^e$ \Comment{$2 \beta_1 \alpha^3 n^3$}
 \State $Y_{xy}^e := A_{zx}^{\overline{e}} X_{zy}^e$ \Comment{$2 \beta_1 \alpha^3 n^3$}
 \State $R_{xy}^e := R_{xy}^e + X_f^e Y_{xy}^f$ \Comment{$2 \beta_1^2 \alpha^2 n^2$}
\State $Y_{xy}^e := X_f^e A_{xy}^f$ \Comment{$2 \beta_1^2 \alpha^2 n^2$}
\State $R_{xy}^e := R_{xy}^e + Y_{zx}^{\overline{e}} X_{zy}^e$  \Comment{$2 \beta_1 \alpha^3 n^3$}
%R calculation
\State $X_{xy}^e := V_{xz} R_{zy}^e$ \Comment{$ \beta_1 \gamma \alpha^2 n^2$}
\State $R_{xy}^e := A_{xy}^e - R_{xy}^e - Y_{xz}^e X_{zy}^e $ \Comment{$2 \beta_1 \alpha^3 n^3$}
 \State $Y_{xy}^e := A_{xz}^e X_{zy}^e$ \Comment{$2 \beta_1 \alpha^3 n^3$}
 \State $R_{xy}^e := R_{xy}^e + X_f^e Y_{xy}^f$ \Comment{$2 \beta_1^2 \alpha^2 n^2$}
 \State $Y_{xy}^e := X_{xy}^e + \Omega_e \nabla_f^e X_{xy}^f$ \Comment{$2 \beta_1^2 \alpha^2 n^2$}
 \State $R_{xy}^e := R_{xy}^e - A_{xz}^e Y_{zy}^e$ \Comment{$2 \beta_1 \alpha^3 n^3$}
\EndFunction
\end{algorithmic}
\end{algorithm}

%P5.11 - Decomposition of T
In terms of half-transformed Green's functions,
\begin{equation}
 G_{ax}^{\underline{a}} = \frac{\phi_{ax}^*}{\omega_{\underline{a}} - \epsilon_a}, \ \ \ G_{xi}^{\underline{i}} = \frac{\phi_{ix}}{\omega_{\underline{i}} - \epsilon_i},
\end{equation}
 Eqs. (\ref{quadrature1}), (\ref{quadrature2}), (\ref{closure}), (\ref{T_factor}), and (\ref{A_ansatz}) reduce $T_{ij}^{ab}$ to
\begin{align} \label{Tsplit}
 T_{ij}^{ab} &\approx - G_{ax}^{\underline{a}} G_{xi}^{\underline{i}} \Omega_{\underline{a}\underline{i}}^e U_{xy}^{ef} 
 \Omega_{\underline{b}\underline{j}}^f G_{by}^{\underline{b}} G_{yj}^{\underline{j}} , \notag \\
 U_{xy}^{ef} &= L_{xz}^{e\overline{g}} \Omega_g V_{zw} L_{yw}^{fg} , \notag \\
 L_{xy}^{ef} &= \delta_f^e \delta_{xy} + \Omega_g \Delta_e^{fg} V_{xz} A_{zy}^{\overline{g}} .
\end{align}
This bears some resemblance to the approximate THC-CCSD form \cite{THC_CCSD} of $T_{ij}^{ab}$.
$U_{xy}^{ef}$ appears in the BRPA analog of Eq. (\ref{fast_sigma0}),
\begin{align}
 \sigma_{xy} &\approx F_{xz}^{\underline{i}} G_{zy}^{\underline{i}} - G_{xz}^{\underline{a}} F_{zy}^{\underline{a}} , \notag \\
 F_{xy}^{\underline{i}} &= \Omega_{\underline{a}\underline{i}}^e G_{xy}^{\underline{a}} W_{yx}^e +
  \Omega_{\underline{a} \underline{j}}^e  \Omega_{\underline{b} \underline{i}}^f G_{xz}^{\underline{a}}
  U_{zy}^{ef} ( \Xi_{zyx}^{\underline{j}\underline{b}} + J_{zy}^{\underline{j}\underline{b}} ) , \notag \\
  F_{xy}^{\underline{a}} &= \Omega_{\underline{a}\underline{i}}^e G_{xy}^{\underline{i}} ( W_{xy}^e + v_x^e ) + 
 \Omega_{\underline{a} \underline{j}}^e  \Omega_{\underline{b} \underline{i}}^f G_{zy}^{\underline{i}}
 U_{xz}^{ef} \Xi_{xzy}^{\underline{j}\underline{b}} , \notag \\
 \Xi_{xyz}^{\underline{i}\underline{a}} &= G_{xw}^{\underline{i}} G_{wy}^{\underline{a}} V_{wz} , \notag \\
 J_{xy}^{\underline{i}\underline{a}} &= G_{xz}^{\underline{i}} f_{zw}^0 G_{wy}^{\underline{a}} , \notag \\
 W_{xy}^e &= \Omega_e V_{xz} A_{zw}^{\overline{e}} V_{wy} , \notag \\
 v_x^e &= U_{xy}^{ef} \Omega_{\underline{a} \underline{i}}^f J_{yy}^{\underline{i}\underline{a}}. \label{fast_sigma}
\end{align}
$\sigma_{xy}$ is calculated by Algorithm \ref{outBRPAf} using $\mathcal{O}(\alpha^3 \beta_2 \beta_3 (\beta_1 + \gamma) n^3)$ operations 
 and $\mathcal{O}(\alpha^2 ( \beta_1 + \beta_2 + \beta_3 ) n^2)$ memory.
A fast BRPA calculation of $\sigma_{xy}$ is a factor of $\mathcal{O}( \beta_1^{-1} \beta_2 \beta_3)$ more expensive than
 a fast MP2 calculation of $\sigma_{xy}^0$ in the $\gamma \gg \beta_m$ regime.
The calculations in Eq. (\ref{fast_sigma}) are especially difficult
 when memory limitations prohibit concurrent storage of $U_{xy}^{ef}$, $J_{xy}^{\underline{i}\underline{a}}$, and $\Xi_{xyz}^{\underline{i}\underline{a}}$.
Full evaluation of $\Xi_{xyz}^{\underline{i}\underline{a}}$ requires $\mathcal{O}(\alpha^3 \beta_2 \beta_3 \gamma n^3)$ operations,
 and without concurrent storage it must be evaluated twice overall.
Without the simple primary and auxiliary basis structure from Sec. \ref{basis_sets},
 pseudocode that is equivalent to Algorithm \ref{outBRPAf} will become significantly longer and more difficult to design.

%A6 - Fast BRPA outer loop
\begin{algorithm}[h!]
\caption{Fast BRPA outer loop\label{outBRPAf}}
\begin{algorithmic}[1]
\Function {outBRPAf}{$\omega_e , \Omega_e , \nabla_f^e , \Omega_{\underline{a}\underline{i}}^e , f_{xy}^0 , V_{xy} , A_{xy}^e , G_{xy}^{\underline{p}} ; \sigma_{xy}$}
\State $\begin{aligned}[t] \mathbf{workspace} : \{ & U^e_f , W_f^e, X_f^e, Y^e_f, Z_f^e, X_{\underline{a}\underline{i}}^e, v_x^e, X_x^{ef}, \\
& X_x^{\underline{a}\underline{i}}, X_x^{e\underline{p}}, Y_x^{e\underline{p}}, W_{xy}^e, X_{xy}^e, X_{xy}^{\underline{i}}, F_{xy}^{\underline{p}} \} \end{aligned}$
\State $F_{xy}^{\underline{p}} := 0$
\State $X_f^e := (1-\delta_f^e)/(\omega_e - \omega_f + \delta_f^e)$
\State $Y_f^e := \Omega_g X_e^{\overline{g}} X_f^g$
%Precalculate J
\State $X_{xy}^{\underline{i}} := G_{xz}^{\underline{i}} f_{zy}^0$
%Calculate W & precalculate U
 \State $X_{xy}^e := \Omega_e V_{xz} A_{zy}^{\overline{e}}$
 \State $W_{xy}^e := V_{yz} X_{xz}^e$
%Calculate v
\State $X_x^{\underline{a}\underline{i}} := X_{xy}^{\underline{i}} G_{yx}^{\underline{a}}$
\State $v_x^e := \Omega_{\underline{a}\underline{i}}^e X_x^{\underline{a}\underline{i}}$
\State $X_x^{ef} := X_{yx}^e v_y^f$
\State $v_x^e := v_x^e + X_f^e (X_x^{ff}  - X_x^{fe}) - \nabla_f^e X_x^{ef}$
\State $v_x^e := \Omega_e V_{xy} v_y^{\overline{e}}$
\State $X_x^{ef} := X_{xy}^e v_y^f$
\State $v_x^e := v_x^e + X_e^f (X_x^{ef}  + X_x^{fe}) - \nabla_e^f X_x^{ff}$
%Calculate F
\ForAll{$x$}
 \State $X_y^{e\underline{a}} := \Omega_{\underline{a} \underline{i}}^e G_{xy}^{\underline{i}}$
 \State $X_y^{e\underline{i}} := \Omega_{\underline{a} \underline{i}}^e G_{yx}^{\underline{a}}$
 %direct parts
 \State $F_{xy}^{\underline{a}} := F_{xy}^{\underline{a}} + X_y^{e\underline{a}} ( W_{xy}^e + v_x^e )$
 \State $F_{yx}^{\underline{i}} := F_{yx}^{\underline{i}} + X_y^{e\underline{i}} W_{xy}^e$
 \ForAll{$y$}
%Calculate U_xy
 \State $Z^e_f := X_{xz}^e W_{yz}^f$
 \State $U^e_f := Y_f^e Z_f^e + \Omega_g ( \nabla_e^{\overline{g}} \nabla_f^g Z^{\overline{g}}_g
     - X_e^{\overline{g}} \nabla_f^g Z^e_g - \nabla_e^g X_f^{\overline{g}} Z^g_f)$
 \State $W^e_f := W_{xy}^f - X_g^e Z^f_g$
 \State $U^e_f := U^e_f + \Omega_f (X_e^{\overline{f}} W^f_e - \nabla_e^{\overline{f}} W^f_{\overline{f}} )$
 \State $W^e_f := W_{yx}^f - X_g^e Z^g_f$
 \State $U^e_f := U^e_f + \Omega_e (X_f^{\overline{e}} W_f^e - \nabla_f^{\overline{e}} W^e_{\overline{e}})$
 \State $U^e_{\overline{e}} := U^e_{\overline{e}} + \Omega_e (V_{xy} - X_f^e W_{xy}^f - X_f^{\overline{e}} W_f^e )$
%exchange parts
  \State $X_{\underline{a}\underline{i}}^e := U^e_f \Omega_{\underline{a} \underline{i}}^f$ \Comment{$2 \beta_1^2 \beta_2 \beta_3 \alpha^2 n^2$}
  \State $X_z^{\underline{a}\underline{i}} := G_{zx}^{\underline{a}} G_{yz}^{\underline{i}}$
  \State $X_z^{\underline{a}\underline{i}} := V_{zw} X_w^{\underline{a}\underline{i}}$ \Comment{$\beta_2 \beta_3 \gamma \alpha^3 n^3$}
  \State $Y_z^{e\underline{i}} := X_z^{e\underline{a}} X_z^{\underline{a}\underline{i}}$
  \State $F_{yz}^{\underline{a}} := F_{yz}^{\underline{a}} + X_{\underline{a}\underline{i}}^e Y_z^{e\underline{i}}$ \Comment{$2 \beta_1 \beta_2 \beta_3 \alpha^3 n^3$}
  \State $X_z^{\underline{a}\underline{i}} := G_{zy}^{\underline{a}} G_{xz}^{\underline{i}}$
  \State $X_z^{\underline{a}\underline{i}} := V_{zw} X_w^{\underline{a}\underline{i}}$ \Comment{$\beta_2 \beta_3 \gamma \alpha^3 n^3$}
  \State $X_z^{\underline{a}\underline{i}} := X_z^{\underline{a}\underline{i}} + X_{xw}^{\underline{i}} G_{wy}^{\underline{a}}$
  \State $Y_z^{e\underline{a}} := X_z^{e\underline{i}} X_z^{\underline{a}\underline{i}}$
  \State $F_{zy}^{\underline{i}} := F_{zy}^{\underline{i}} + X_{\underline{a}\underline{i}}^e Y_z^{e\underline{a}}$ \Comment{$2 \beta_1 \beta_2 \beta_3 \alpha^3 n^3$}
 \EndFor
\EndFor
 \State $\sigma_{xy} := F_{xz}^{\underline{i}} G_{zy}^{\underline{i}} - G_{xz}^{\underline{a}} F_{zy}^{\underline{a}}$
\EndFunction
\end{algorithmic}
\end{algorithm}

%S6
\section{Applications\label{applications}}
\vspace{-7pt}

%P6.1 - Basic motivation for semiempirical calculations
Before committing further to its development, it is prudent to test the accuracy of BRPA theory against
 other popular total energy methods and performance of the fast MP2 and BRPA algorithms
 against their conventional counterparts.
Direct use of the algorithms in Sec. \ref{fast_design} is possible for Hamiltonians that are limited
 to a zero-differential-overlap (ZDO) form \cite{ZDO}.
There have been proposals to calculate electron correlation within a semiempirical framework \cite{MNDOC},
 but they remains undeveloped.

%F1 - ZDO H_2 dissociation curves (exact, HF, MP2, B3LYP, BRPA)
\begin{figure*}
\includegraphics{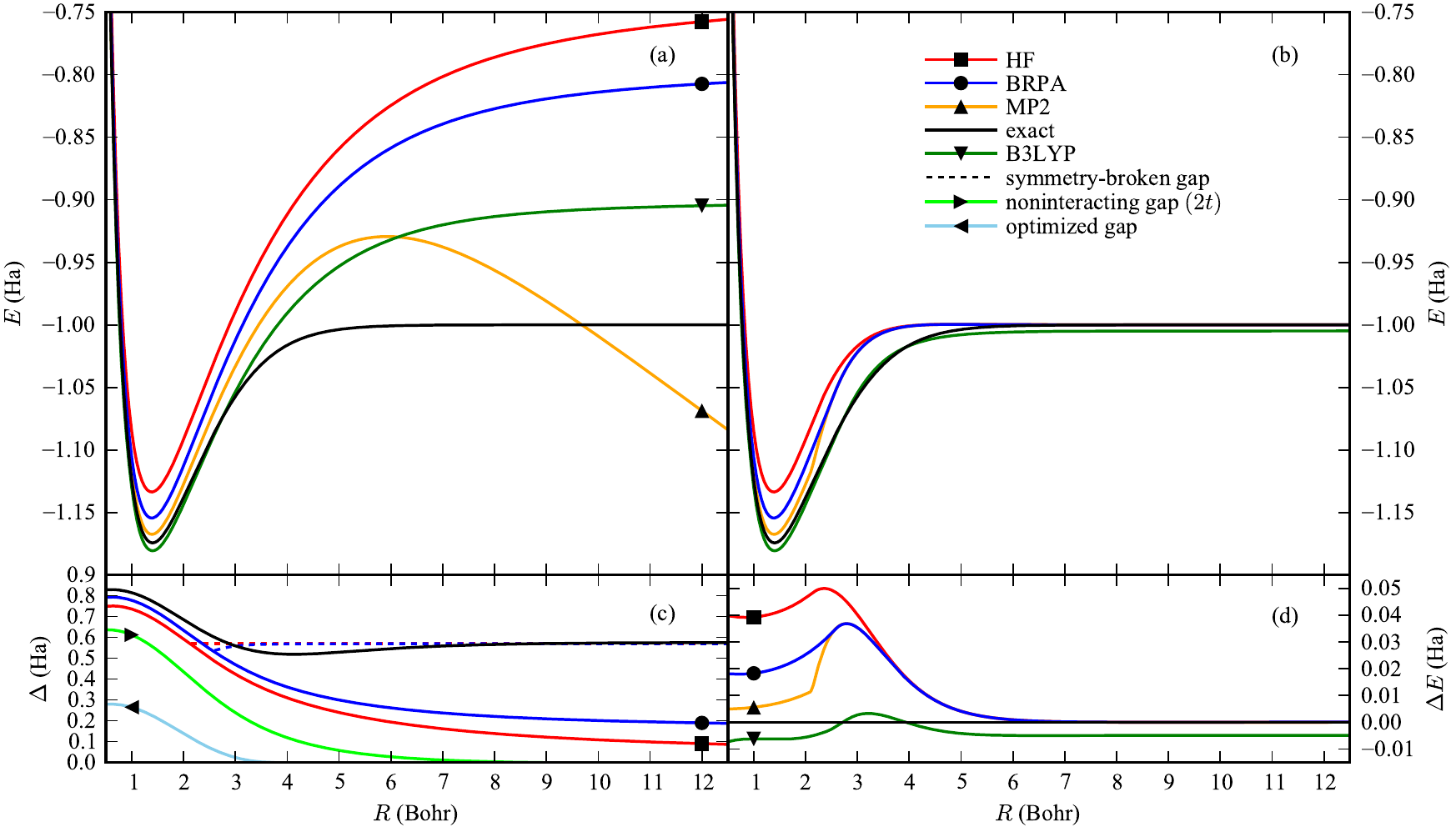}
\caption{\label{H2energy} Total energy, $E$, with symmetry (a) preserved and (b) broken,
 (c) virtual-occupied energy gap, $\Delta$, and (d) error in the symmetry-broken total energy, $\Delta E$, for H$_2$
 as a function of bond length, $R$.
The RPA+SOSEX correlation energy is exact for an optimized gap when $R < 3.7$.}
\end{figure*}

%P6.2 - Overview of the H_n semiempirical model
We consider H$_n$ because it gives us fine control over the number of electrons and
 also because H$_2$ is the simplest two-electron molecule with an internal coordinate.
The ZDO form of H$_2$ is an extended Hubbard model with $\alpha = 2$ and
\begin{equation} \label{H_H2}
 \mathbf{h} = \scriptsize \left[ \begin{array}{cccc} \mu & -t & 0 & 0 \\ -t & \mu & 0 & 0 \\
 0 & 0 & \mu & -t \\ 0 & 0 & -t & \mu \end{array} \right] \normalsize , \ \ \ 
 \mathbf{V} = \scriptsize \left[ \begin{array}{cccc} U & V & U & V \\ V & U & V & U \\
 U & V & U & V \\ V & U & V & U \end{array} \right] \normalsize ,
\end{equation}
 in the notation of Eq. (\ref{aux_H}).
It has a natural pairwise extension to the $n > 2$ case.
Separate H$_2$ and H$_n$ Hamiltonians are fit in Appendix \ref{Hn_fit}.
The H$_2$ version is fit to reproduce exact, HF, and MP2 total energies.
The H$_n$ version is fit to a simple distance-dependent form
 that gives proper asymptotic behavior.
These models have limited transferability, which is a standard caveat of
 semiempirical modeling with simple Hamiltonian forms.

%S6.1
\vspace{-2pt}
\subsection{Accuracy of BRPA on H$_2$}
\vspace{-7pt}

%P6.3 - Simple analytic results in the symmetric case
Symmetric dissociation of H$_2$ is used as a critical test of new
 electron correlation methods \cite{H2_SOSEX,H2_RPA,H2_GW}.
The model in Eq. (\ref{H_H2}) has simple formulas
 for HF, MP2, BRPA, and exact energies,
\begin{align}
 E_{\mathrm{HF}} &= E_0 + 2 (\mu - t) +\tfrac{1}{2} (U + V), \notag \\
 E_{\mathrm{MP2}} &= E_{\mathrm{HF}} - \tfrac{(U - V)^2}{16 t + 8 V} , \ \ \ 
 E_{\mathrm{BRPA}} = E_{\mathrm{HF}} - \tfrac{(U - V)^2}{16 t + 8 U} , \notag \\
 E &= E_{\mathrm{HF}} + 2 t - \sqrt{ 4 t^2  + \tfrac{1}{4}(U-V)^2} .
\end{align}
Dissociation curves are shown in Fig. \ref{H2energy}a.
BRPA is a uniform improvement over HF but has three times the error of
 MP2 or B3LYP at equilibrium.
These methods fail at dissociation, but
 BRPA is stable at twice the B3LYP error as MP2 diverges.

%P6.4 - Choice of reference state
RPA+SOSEX total energies are strongly modulated by the choice of mean field.
As shown in Fig. \ref{H2energy}c, errors in H$_2$ near equilibrium can be corrected by significantly reducing the gap.
This is typical behavior, and the smaller orbital gaps in DFT reference mean fields
 systematically improve total energies \cite{RPArev1} over HF-based calculations \cite{RPA3rd}.
Reducing gaps to reduce errors conflicts with the generalized Koopmans theorem in Eq. (\ref{general_Koopmans})
 that requires the BRPA gap to approximate the physical gap.
On H$_2$, BRPA increases the HF gap towards the exact gap, but some methods
 for calculating excitations such as $GW$ theory decrease the HF gap \cite{H2_GW}.
One possible resolution is to derive an RPA+SOSEX  model where orbital energies
 are not required to approximate electron addition and removal energies,
 such as by modifying $P_{xy}^{\lambda}(\omega)$ in Eq. (\ref{ACFD_RPA})
 with exchange terms \cite{Hubbard_SOSEX}.

%P6.5 - Symmetry-broken case
Dissociation failures can be fixed by enabling symmetry breaking in the reference state.
As shown in Fig. \ref{H2energy}d, the error now peaks at around twice the equilibrium bond length.
MP2 now has uniformly less error than BRPA, but BRPA fixes the derivative discontinuity in the MP2 energy surface.
Symmetry breaking is needed for size consistency at this level of theory
 when considering the dissociation of a closed-shell molecule into open-shell fragments.
It also serves as a simple model of static correlation.
Broken symmetries can be restored with a projection \cite{PHF},
 but the resulting total energy corrections are not size consistent.
An accurate, size-consistent model of electron correlation should be able to repair
 symmetries broken by the reference state accurately but not necessarily exactly.

%P6.6 - Value of a single test case
A single example such as H$_2$ is not enough to determine error statistics
 and compare the average errors of total energy methods.
However, while errors on a system as important and simple as H$_2$ remain large,
 it suffices to discount methods.

%T1 - Precise operations and memory counts
\begin{table*}[!t]
\caption{\label{cost_table}Leading-order, per-iteration floating-point costs and memory footprints of MP2 and the inner and outer loops of BRPA.}
\footnotesize
\begin{tabular}{l r r}
\hline \hline
 Algorithm & Operations & Memory \\
\hline
 \ref{MP2c}. Conventional MP2 & $4 \alpha^3 n^5$ & $2.5 \alpha^2 n^2$ \\
 \ref{inBRPAc}. Conventional Inner BRPA & $4 \alpha^3 n^5$ & $2 \alpha^2 n^3$ \\
 \ref{outBRPAc}. Conventional Outer BRPA & $4 \alpha^3 n^5$ & $2 \alpha^2 n^3$ \\
 \ref{MP2f}. Fast MP2 & $6 \beta_1 \beta_2 \beta_3 \alpha^2 n^2 + 2 \beta_1 ( 2 \beta_2 + 3 \beta_3 + \gamma ) \alpha^3 n^3$ & $\beta_1 \beta_2 \beta_3 + (\beta_1 \beta_2 + \beta_1 \beta_3) \alpha n + (3 \beta_1 + 2 \beta_2 + 2 \beta_3) \alpha^2 n^2$ \\
 \ref{inBRPAf}. Fast Inner BRPA & $2 \beta_1 ( 5 \beta_1 + \gamma ) \alpha^2 n^2  + 12 \beta_1 \alpha^3 n^3$ & $ 2 \beta_1^2 + 4.5 \beta_1 \alpha^2 n^2$ \\
 \ref{outBRPAf}. Fast Outer BRPA & $2 \beta_1^2 \beta_2 \beta_3  \alpha^2 n^2 + 2 (2 \beta_1 + \gamma) \beta_2 \beta_3  \alpha^3 n^3$ &  $2 \beta_1 \beta_2 \beta_3 + (\beta_1^2 + 2 \beta_1 \beta_2 + 2 \beta_1 \beta_3 + \beta_2 \beta_3) \alpha n + ( 3 \beta_1 + 2 \beta_2 + 3 \beta_3 ) \alpha^2 n^2$ \\
\hline \hline
\end{tabular}
\end{table*}

%S6.2
\subsection{Scaling of BRPA on H$_n$}
\vspace{-7pt}

%P6.7 - Calculation details
We benchmark the algorithms in Sec. \ref{fast_design} on H$_n$ rings,
 with the intent to produce representative scaling behavior and not necessarily
 to optimize performance.
To this end, we consider only $n = 4 m + 2$ for a closed shell and a finite orbital gap.
We use an HF mean field in all calculations to enable the use of common quadratures.
The algorithm runtimes are comparable to per-iteration costs
 and a full self-consistent calculation will have number of iterations as an additional cost multiplier.

%P6.8 - Quadrature prefactors
The HF orbital gap decreases as $\Delta \approx 2.7 n^{-0.83}$.
Using the quadrature formulas in Appendix \ref{quadrature_section}
 with errors that are set to $\varepsilon_Q = 10^{-5}$,
 the quadrature sizes are observed to increase as
\begin{align}
 \beta_1 &\approx 1.7 + 2.1 \ln n , \notag \\
 \beta_2 &\approx -0.7 + 4.1 \ln n , \notag \\
 \beta_3 &\approx 3.3 + 4.1 \ln n .
\end{align}
These are not asymptotic scalings.
When the orbital gap falls below a numerical low-energy cutoff,
 we will add an artificial gap comparable to this cutoff
 that limits quadrature size.

%P6.9 - Implementation details & summation prefactor
Benchmarks are measured on a \textsc{c} implementation \cite{supplement}.
The HF Hamiltonian matrix is diagonalized with the \textsc{dsyev} routine in \textsc{lapack} \cite{LAPACK}.
Tensors are contracted with the real \textsc{dgemm} and complex \textsc{zgemm} routines in \textsc{blas} \cite{BLAS}.
A conversion factor from
 runtimes to operation counts is determined by timing
 the $2 n^3$ operations of $n$-by-$n$ matrix-matrix multiplications in \textsc{dgemm} calls for large $n$.
Conventional algorithms use real arithmetic,
 while fast algorithms use complex arithmetic.
The increased cost of complex arithmetic is mitigated by using symmetry \cite{symmetry_note}.
$V_{xy} \rho_{y}$ is summed with \textsc{fftw} \cite{FFTW}.
For regular behavior of $\gamma$, we further restrict H$_n$ to $n = 2^p - 2$
 and embed $V_{xy} \rho_y$ in a cyclic convolution of size $2^{p+1}$.
We observe an \textsc{fftw} scaling of
\begin{equation}
 \gamma \approx 23 + 13 \ln n .
\end{equation}
Other fast summation methods \cite{fast_kernel_sum} avoid a $\ln n$ prefactor.

\begin{figure}[!b]
\includegraphics{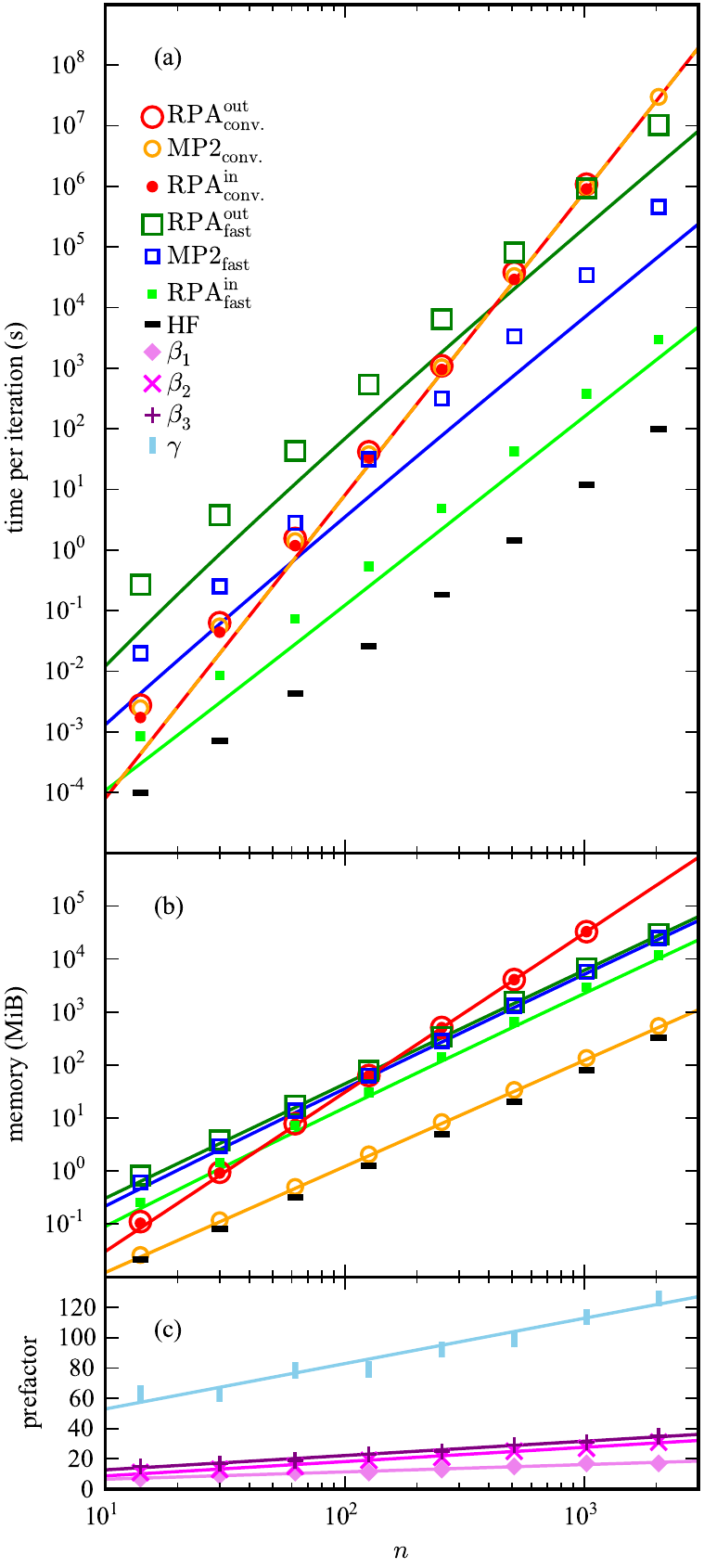}
\caption{\label{bench} The (a) operation, (b) memory, and (c) prefactor costs of H$_n$ on a single core of an Intel Xeon X5650.
Points are measured costs, and the equivalently colored lines are the model costs \cite{cost_note}.
The model conversion factor between operation count and runtime is 10 Gflops.
}
\end{figure}

%P6.10 - Comparison of measured and modeled costs
Benchmark results are shown in Fig. \ref{bench}.
The runtimes are compared to the model estimates summarized in Table \ref{cost_table}.
The agreement is good, and all discrepancies can be rationalized.
The \textsc{inBRPAf} cost is increased $50 \%$ by complex arithmetic \cite{symmetry_note}.
The excess cost of \textsc{MP2f} is caused by \textsc{zgemm} calls with small $\mathcal{O}(\beta_1)$ matrix dimensions.
The excess cost of \textsc{outBRPAf} is caused by subleading-order terms with large prefactors.
Fast and conventional calculations agree to $10^{-7}$ Ha in $\sigma_{xy}$.

%P6.11 - What these results mean for "accurate" calculations
In the cost model, the crossover between conventional and fast runtimes
 occurs at $n=62$ for MP2 and $n=422$ for BRPA.
As all algorithms have an $\mathcal{O}(\alpha^3)$ scaling,
 larger basis sets will not shift the crossover.
Other prefactors will but are indicative of realistic values
 as quadrature sizes are reported \cite{oldfastRPA,VASP_RPA} from $16$ to $40$
 and the FFT is a standard fast summation method.
 
%F2 - H_n scaling plot

%S7
\section{Discussion\label{discussion}}
\vspace{-7pt}

%P7.1 - What is needed to implement this in electronic structure codes?
A natural continuation of the research in this paper is the
 extension of the algorithms in Sec. \ref{fast_design} for implementation in established electronic structure codes.
For a Gaussian-orbital code, this means nontrivial overlap matrices
 between primary basis functions and between pairs of primary basis functions and auxiliary basis functions.
It needs an implementation of THC or some other decomposition of $V_{rs}^{pq}$ with a sparse vertex tensor
 and a Gaussian-compatible FMM \cite{CFMM}.
This combination of features is not yet available in any code.
For a planewave-pseudopotential code, algorithms for periodic systems need to be developed.
Suitable FFT and grid operations for efficient $V_{rs}^{pq}$ decomposition are widely available in these codes.

%P7.2 - Other future research directions
An orthogonal direction for continued research is further
 development of basic algorithms and numerical analysis that improve cubic-scaling correlation models.
One such direction is analysis of the slow basis set convergence that
 is a universal detriment to non-DFT electron correlation models.
Another is the development of more accurate correlation models
 that are restricted to $\mathcal{O}(\alpha^3 n^3)$ operations and $\mathcal{O}(\alpha^2 n^2)$ memory.
These two directions are discussed in more detail below.

\vspace{-2pt}
\subsection{The operator approximation problem\label{op_approx}}
\vspace{-7pt}

%P7.3 - alpha determined by orbital approximation
The $\alpha$ prefactor depends on the choice of basis set.
Basis construction is widely considered as a function approximation problem of electron orbitals.
It is usually focused on occupied orbitals, their response to perturbations, and low-lying virtual orbitals.
Larger $\alpha$ values improve their approximation,
 while smaller $\alpha$ values reduce costs.
The number of virtual orbitals defined by the basis set is also determined by $\alpha$.
 
%P7.4 - Existing approaches to improving virtual orbital sum convergence
Slow basis set convergence manifests in the virtual orbital summations
 in MP2 and related calculations.
Convergence is often accelerated with basis set extrapolation
 to avoid large $\alpha$ values \cite{extrapolation}.
In wavefunction-based methods, slow convergence is attributed to
 electron-electron cusps in the wavefunction \cite{cusp}
 that are corrected by R12/F12 \cite{R12} or Jastrow \cite{Jastrow} methods.
In the fast algorithms in Sec. \ref{fast_design},
 there are only operators and their finite-basis errors
 instead of summations over virtual orbitals or wavefunction cusps.
In the case of a spinless nonrelativistic Hamiltonian, the
 BRPA operator variables asymptote to
\begin{equation}
 G(\vec{x},\vec{y},\omega) \rightarrow \frac{-1}{2 \pi |\vec{x}-\vec{y}|}, \ \ \
 A(\vec{x},\vec{y},\omega) \rightarrow \frac{-\rho(\vec{x},\vec{y})}{2 \pi |\vec{x}-\vec{y}|},
\end{equation}
 as $|\vec{x}-\vec{y}| \rightarrow 0$.
These singularities converges slowly with basis set size
 and are better approximated by functions of $|\vec{x}-\vec{y}|$.

%P7.5 - Interpretation as an operator approximation problem
The efficient and accurate representation of $A(\mathbf{x},\mathbf{y},\omega)$ and $G(\mathbf{x},\mathbf{y},\omega)$
 is an operator approximation problem.
$G(\mathbf{x},\mathbf{y},\omega)$ is the simplest case where we seek to satisfy
\begin{equation}
 \int d \mathbf{z} [ \omega \delta(\mathbf{x}-\mathbf{z}) - f(\mathbf{x},\mathbf{z}) ] G(\mathbf{z},\mathbf{y},\omega)  \approx \delta(\mathbf{x} - \mathbf{y}) .
\end{equation}
The conventional approach is to approximate $G(\mathbf{x},\mathbf{y},\omega)$
 with a sum of pairwise products of basis \textit{functions}.
The possibilities beyond this include basis \textit{operators} that are added linearly or as pairwise products
 and the direct modeling of singularities.
An operator-based approach can also use spatial localization \cite{linear_scaling}
 that does not manifest in molecular orbitals.
These ideas form a distinct alternative to orbital function approximation.

\subsection{The cost-restricted electron correlation problem}
\vspace{-7pt}

%P7.6 - The many faces of RPA
The BRPA model developed in this paper is in the class of RPA models,
 but it is derived from BCCD theory.
Historically, the mathematical structure of RPA has emerged from multiple theories: %chronological refs:
 resummation of diagrammatic perturbation theory \cite{RPA_egas}, %1957
 boson approximation of electron-hole excitation operators \cite{RPA_transform}, %1957
 equations-of-motion for ground state annihilation operators \cite{RPA_EOM}, %1960
 integration of response functions over interaction strength \cite{RPA_AC}, %1964
 integration of Slater determinants over generator variables  \cite{RPA_GCM}, %1964
 $1/N$ expansion in $N$ interacting copies of the Hilbert space \cite{RPA_largeN}, %1965
 total energy functionals of the many-body Green's function \cite{GW}, %1965
 and an adaptation of interaction strength integration to DFT \cite{RPA_DFT}. %1977
This diverse set of theoretical ideas is responsible for the large number of modern RPA variants
 \cite{RPArev1,RPArev2,RPArev3}.
Much like DFT, there is no rigorous delineation on what can be incorporated into an RPA model of electron correlation energy.

%P7.7 - Method development delineated by scaling
The results of this paper suggest a delineation of models based on computational complexity:
 a cost strictly bound by a scaling of $\mathcal{O}(n^3)$ operations and $\mathcal{O}(n^2)$ memory
 with a limit on the form of allowed approximations.
Other approximations such as orbital localization \cite{local_MP2} and stochastic sampling \cite{MC_MP2,MC2_MP2}
 can further reduce the scaling of MP2-like methods but introduce localization length
 and number of samples as complications.
A critical question is whether or not these cost restrictions
 are sufficient to enable method development based on the direct approximation of a quantum state as in CC theory
 or if it is to be confined to the indirect and often empirical development that is characteristic of modern DFT \cite{Jacobs_ladder}.

%S8
\vspace{-2pt}
\section{Conclusions\label{conclusions}}
\vspace{-7pt}

%P8.1 - Summary of results
The results of this paper cross an important threshold.
For the first time, the computational cost of an electron correlation model with a direct connection
 to an underlying wavefunction has been reduced to the canonical cost of an SCF calculation,
 $\mathcal{O}(n^3)$ operations and $\mathcal{O}(n^2)$ memory,
  without localization or stochastic approximations.
This includes MP2 theory and an RPA+SOSEX approximation of BCCD theory.
The new ideas used to achieve this result may also contribute to reducing the cost of more accurate
 electron correlation models.

%P8.2 - Cubic-scaling correlation as a distinct paradigm
Electron correlation models that require $\mathcal{O}(n^3)$ operations and $\mathcal{O}(n^2)$ memory
 could constitute a distinct paradigm with continued development.
They are a compromise between the low cost of DFT and the high accuracy of CC theory.
For this compromise to be worthwhile,
 we need to explore the balance between cost and accuracy
 to identify its fundamental limits.

\acknowledgments
\vspace{-7pt}
I thank Jay Sau, Norm Tubman, Jeff Hammond, Andrew Baczewski,
 Rick Muller, and Toby Jacobson for discussions.
I thank Andrew Baczewski for checking the mathematics and proofreading the manuscript.
This work was supported by the Laboratory Directed Research and Development program at Sandia National Laboratories.
Sandia National Laboratories is a multi-program laboratory managed and  
operated by Sandia Corporation, a wholly owned subsidiary of Lockheed 
Martin Corporation, for the U.S. Department of Energy's National  
Nuclear Security Administration under contract DE-AC04-94AL85000.

\appendix

%SA
\section{Numerical quadrature\label{quadrature_section}}
\vspace{-7pt}

%PA.1 - Problem statement
An implementation of the fast MP2 and BRPA algorithms requires numerical quadratures
 for the integrals appearing in Eqs. (\ref{quadrature1}) and (\ref{quadrature2}).
We restate these integrals and quadratures in a more general notation with explicit summations,
\begin{align} \label{quadrature}
 \frac{1}{a + a'} &= \int_{-i \infty}^{i\infty} \frac{dx}{2\pi i} \frac{1}{(a-x)(a'+x)} \notag \\
 &\approx \sum_{i=1}^{n_x} \frac{w_i}{(a - x_i)(a' + x_i)} , \ \ \ a,a' \in A , \notag \\
 \frac{\theta(b,B) \theta(c,C)}{b - c + x} &=  \oint_{Y} \frac{d y}{2 \pi i} \oint_{Z} \frac{d z}{2 \pi i} \frac{1}{(b - y)(c - z) (y - z + x) } \notag \\
 &\approx \sum_{i=1}^{n_y} \sum_{j=1}^{n_z} \frac{W_{ij}(x)}{(b - y_i)(c - z_j)} , \ \ \ b,c \in B \cup C , \notag \\
 \theta(s,S) &= 1 \ \ \ \forall s \in S, \ \ \ \theta(s,S) = 0 \ \ \ \forall s \not\in S .
\end{align}
$A$ is the set of orbital transition energies ($\epsilon_a - \epsilon_i$),
 $B$ is the set of virtual orbital energies ($\epsilon_a$),
 $C$ is the set of occupied orbital energies ($\epsilon_i$),
 $X$ is the set of imaginary reals,
 and $Y$ and $Z$ are closed counterclockwise contours that satisfy
\begin{align}
 B &\subset \mathrm{in}(Y), \ \ \  (B+X) \cap \mathrm{in}(Z) = \varnothing, \notag \\
 C &\subset \mathrm{in}(Z), \ \ \ (C-X) \cap \mathrm{in}(Y) = \varnothing ,
\end{align}
 for interior, $\mathrm{in}(S) = \{t s + (1-t) s' : t \in [0,1], s,s' \in S\}$,
 and elemental arithmetic, $S \pm S'=\{s \pm s':s \in S, s' \in S'\}$, set operations.
We assume $A = B - C$ and $a > 0 \ \forall a \in A$.

%PA.2 - Construction from simple quadratures
Eq. (\ref{quadrature}) is assembled from more elementary quadratures.
We split the first integral into two terms with partial fractions,
\begin{equation}
 \frac{1}{a + a'} = \frac{1}{a + a'} \int_{-i \infty}^{i\infty} \frac{dx}{2\pi i} \left[ \frac{1}{a-x}  - \frac{1}{-a'-x} \right] ,
\end{equation}
 and fit quadrature to its two terms on a merged domain,
\begin{equation} \label{qA}
 \theta(a,A) - \frac{1}{2} =  \int_{-i \infty}^{i\infty} \frac{dx}{2\pi i} \frac{1}{a-x} \approx \sum_{i=1}^{n_x} \frac{w_i}{a - x_i}, \ \ \ a \in A' ,
\end{equation}
 for $A' = A \cup \{-a: a \in A\}$.
For the second integral, we split with partial fractions after performing one of the integrations,
\begin{align}
 \frac{\theta(b,B)\theta(c,C)}{b - c + x} &= \frac{\theta(c,C)}{b - c + x} \oint_{Y} \frac{d y}{2 \pi i} \left[ \frac{1}{b-y} - \frac{1}{c-x-y} \right] , \notag \\
 &= \frac{\theta(b,B)}{b - c + x} \oint_{Z} \frac{d z}{2 \pi i} \left[ \frac{1}{c-z} - \frac{1}{b+x-z} \right] , \notag
\end{align}
 and again fit quadratures to each contour on a merged domain,
\begin{align} \label{qBC}
 \theta(b,B) &= \oint_{Y} \frac{d y}{2 \pi i} \frac{1}{b-y} \approx \sum_{i=1}^{n_y} \frac{u_i}{b - y_i}, \ \ \ b \in B' , \notag \\
 \theta(c,C) &= \oint_{Z} \frac{d z}{2 \pi i} \frac{1}{c-z} \approx \sum_{i=1}^{n_z} \frac{v_i}{c - z_i}, \ \ \ c \in C' ,
\end{align}
 for $B' = B\cup(C-X)$ and $C' =  C\cup(B+X)$.
The fits combine to produce the quadrature weight, $W_{ij}(x) = u_i v_j/ (y_i - z_j + x)$.
We assume that the reassembly of these quadratures is stable and leave further error analysis to future work.

%PA.3 - Reduction & 1-parameter solution of the first quadrature
We approximate Eq. (\ref{qA}) with a 1-parameter quadrature,
\begin{equation} \label{qsgn}
  \frac{1}{2} \mathrm{sgn}(a') \approx \sum_{i=1}^{2n+1} \frac{w'_i}{a' - x'_i}, \ \ \ a' \in [ -1, -k] \cup [ k, 1 ] , 
\end{equation}
 for $0 < k < 1$ and map back to the original quadrature with
\begin{equation}
 \scriptsize \begin{bmatrix} w_i \\[-2.5pt] x_i \\[-2.5pt] a \end{bmatrix} \normalsize =  \max(A)  \scriptsize \begin{bmatrix} w'_i \\[-2.5pt] x'_i \\[-2.5pt] a' \end{bmatrix} \normalsize, \ \ \  k = \tfrac{\min(A)}{\max(A)}, \ \ \ n_x = 2 n + 1.
\end{equation}
Zolotarev's rational approximation of the sign function using Jacobi elliptic functions
 minimizes the maximum error \cite{Zolotarev},
\begin{align} \label{Zsign}
 \varepsilon &= \tfrac{1}{2} \tfrac{r(k) - r(\kappa)}{r(k) + r(\kappa)} , \ \ \ 
 r(x) = \tfrac{1}{x} \prod_{m=1}^n \tfrac{x^2 + f^2_{2m-1}}{x^2 + f^2_{2m}} , \notag \\
 f_m &= k \tfrac{\mathrm{sn}( m \theta, k')}{\mathrm{cn}( m \theta, k')}, \ \ \
w'_1 = \tfrac{1}{r(k) + r(\kappa)} \prod_{m=1}^n \tfrac{f_{2m-1}^2}{f_{2m}^2}, \notag \\
 w'_{2m} &= w'_{2m+1} = \tfrac{1}{2} \tfrac{1}{r(k) + r(\kappa)} \prod_{p = 1}^n \tfrac{f_{2m}^2 - f_{2p-1}^2}{f_{2m}^2 - f_{2p}^2 ( 1 - \delta_{pm})} , 
 \notag \\
 x'_1 &= 0, \ \ \ x'_{2m} = -x'_{2m+1} = i f_{2m} , \notag \\
 k' &= \sqrt{1 - k^2}, \ \ \  \kappa = \tfrac{k}{\mathrm{dn}( \theta,k' )} , \ \ \ \theta = \tfrac{K'(k)}{2n+1}.
\end{align}
$\varepsilon$ is the maximum error, which decays exponentially in $n$ with
 an exponent of $2 \pi K(k) / K'(k)$
 that asymptotes to $\pi^2/\ln(4/k)$ in the $k \rightarrow 0$ limit \cite{sgn_decay}.
$\mathrm{sn}(u,k)$, $\mathrm{cn}(u,k)$, and $\mathrm{dn}(u,k)$ are Jacobi elliptic functions.
$K(k)$ and $iK'(k)$ are their quarter periods.

%PA.4 - Mapping problem for the second quadrature
We approximate Eq. (\ref{qBC}) with a 1-parameter quadrature,
\begin{equation}\label{qstep}
 \mathrm{sgn}(\mathrm{Re}(b')) \approx \sum_{i=1}^{2 n + 1} \frac{u'_i}{b' - y'_i}, \ \ \ b' \in X \cup [ \lambda, 1],
\end{equation}
 for $0 < \lambda < 1$ and map back to the original quadratures with
\begin{align}
 &\scriptsize \begin{bmatrix} u_i \\[-2.5pt] y_i - \max(C) \\[-2.5pt] b - \max(C) \end{bmatrix} \normalsize
 = \left[\max(B) - \max(C)\right] \scriptsize \begin{bmatrix} u'_i \\[-2.5pt] y'_i \\[-2.5pt] b' \end{bmatrix} \normalsize , \notag \\
 &\lambda = \tfrac{\min(B) - \max(C)}{\max(B) - \max(C)}, \ \ \ n_y = 2 n + 1, \ \ \ \mathrm{or} \notag \\
 &\scriptsize \begin{bmatrix} v_i \\[-2.5pt] z_i - \min(B) \\[-2.5pt] c - \min(B) \end{bmatrix} \normalsize
 = \left[\min(C) - \min(B)\right] \scriptsize \begin{bmatrix} u'_i \\[-2.5pt] y'_i \\[-2.5pt] b' \end{bmatrix} \normalsize , \notag \\
 &\lambda =  \tfrac{\max(C) - \min(B)}{\min(C) - \min(B)} , \ \ \ n_z = 2 n + 1.
\end{align}
The interiors of $B+X$ and $C-X$ are omitted from Eq. (\ref{qstep})
 by the maximum modulus principle.
We map from Eq. (\ref{qsgn}) to Eq. (\ref{qstep}) with
 a shift of $\mathrm{sgn}(a')$ by $1/2$ and a transformation,
\begin{equation}
 a'= \tfrac{(1+ k) (b')^2 - 2 k}{2 - (1+ k) (b')^2}, \ \ \
 k = \tfrac{\lambda^{2} }{ 2 - \lambda^{2} + 2 \sqrt{1 - \lambda^2} } .
\end{equation}
The rational transformation creates a reflection of the original contour
 about the imaginary axis and doubles the number of poles.
We remove the new contour and its poles leaving
\begin{align}
 u'_i = \tfrac{w'_i}{y'_i} \tfrac{1 - k}{(1+k)(1+x'_i)^2} , \ \ \
  y'_i = \sqrt{2 \tfrac{k + x'_i}{(1+k)(1+x'_i)}} .
\end{align}
Empirically, we observe that the maximum error is twice $\varepsilon$ in Eq. (\ref{Zsign}).
This approximant does not minimize the maximum error exactly,
 but we conjecture that the exponential decay rate of the error with $n$ is optimal.

%SB
\section{Semiempirical H$_n$ model\label{Hn_fit}}
\vspace{-7pt}

%PB.1 - Basic model and matrix elements
To solve the semiempirical H$_2$ model in Eq. (\ref{H_H2}), we use
 a family of reference orbitals parameterized by $\theta$,
\begin{align}
 \boldsymbol{\phi}_a &= \scriptsize \begin{bmatrix}
 \sin \theta \\[-2.5pt]
 - \cos \theta \\[-2.5pt]
 0 \\[-2.5pt]
 0 \end{bmatrix} \normalsize, \ \ \
 \boldsymbol{\phi}_b = \scriptsize \begin{bmatrix}
 0 \\[-2.5pt]
 0 \\[-2.5pt]
 -\cos \theta \\[-2.5pt]
 \sin \theta \end{bmatrix} \normalsize, \notag \\
 \boldsymbol{\phi}_i &= \scriptsize \begin{bmatrix}
 \cos \theta \\[-2.5pt]
 \sin \theta \\[-2.5pt]
 0 \\[-2.5pt]
 0 \end{bmatrix} \normalsize, \ \ \
 \boldsymbol{\phi}_j = \scriptsize \begin{bmatrix}
 0 \\[-2.5pt]
 0 \\[-2.5pt]
 \sin \theta \\[-2.5pt]
 \cos \theta \end{bmatrix} \normalsize,
\end{align}
 where $\{i,j\}$ label the two occupied orbitals and $\{a,b\}$ label the two virtual orbitals.
The nonzero matrix elements are
\begin{align}
 h_i^i &= h_j^j = \mu - t \sin 2 \theta , \notag \\
 h_a^a &= h_b^b = \mu + t \sin 2 \theta , \ \ \ h_a^i = h_b^j = t \cos 2 \theta , \notag \\
 V_{pp}^{pp} &= V_{aj}^{aj} = V_{bi}^{bi} = U - \tfrac{1}{2} (U - V) (\sin 2 \theta)^2 , \notag \\
 V_{ai}^{ai} &= V_{bj}^{bj} = V_{ab}^{ab} = V_{ij}^{ij} = V + \tfrac{1}{2} (U - V) (\sin 2 \theta)^2, \notag \\
 V_{aa}^{ii}  &= V_{bb}^{jj} = -V_{ab}^{ij} = \tfrac{1}{2} (U-V) (\sin 2 \theta)^2 , \notag \\
 V_{ab}^{ib} &= V_{ai}^{ii} = V_{ba}^{ja} = V_{bj}^{jj} = -V_{aa}^{ia} = -V_{aj}^{ij}  \notag \\
 &= -V_{bb}^{jb} = -V_{bi}^{ji} = \tfrac{1}{4} (U - V) \sin 4 \theta ,
\end{align}
 with symmetric extensions for $_p^q \Leftrightarrow \, \! _q^p$ and $_{rs}^{pq} \Leftrightarrow \, \! _{sr}^{qp}$.

%PB.2 - Basic equations: HF, BCCD, & BRPA
The BCCD total energy in terms of $T = T_{ij}^{ab}$ is
\begin{align} \label{E_H2}
 E &= E_0 + 2 (\mu - t \sin 2 \theta) + V \notag \\ 
 & \ \ \ + \tfrac{1}{2} (U-V) (1-T) (\sin 2 \theta)^2  .
\end{align}
The mean field equations are an orbital energy difference, $\Delta$, and a coupling
 between occupied and virtual orbitals,
\begin{subequations}\begin{align}
 \Delta &= 2 t \sin 2 \theta +  U - (U-V) (1-T) (\sin 2 \theta)^2 , \label{H2_gap} \\
 0 &= [ t (1+T) - \tfrac{1}{2} (U - V) (1-T) \sin 2 \theta ] \cos 2 \theta .
\end{align}\end{subequations}
These are the complete HF equations if $T=0$.
There are up to two solutions: a nonsymmetric state and a symmetric state for $\sin 2 \theta = 1$.
In BCCD theory, $T$ is the solution to Eq. (\ref{BCCD_double}),
\begin{equation}
 0 = 2 ( \Delta - U ) T - \tfrac{1}{2} (U-V) (1 - 4 T + 3 T^2)  (\sin 2 \theta)^2 .
\end{equation}
The RPA Riccati equation in Eq. (\ref{BRPA_double}) similarly reduces to
\begin{equation}
  0 = 2 \Delta T - \tfrac{1}{2}(U-V) (1 - 4 T + 4 T^2) (\sin 2 \theta)^2 ,
\end{equation}
 which simplifies the 4-by-4 matrix equation for $T_{ij}^{ab}$ to a direct condition on $T$
 by exploiting matrix symmetries.

%PB.3 - T solutions in the symmetric case
For $\theta = \tfrac{1}{4} \pi$, the quadratic formula is used to solve for $T$,
\begin{align}
 T(\mathrm{exact}) &= \sqrt{\left[ \tfrac{4t}{U-V} \right]^2 + 1} - \tfrac{4t}{U-V} \le 1, \notag \\
 T(\mathrm{RPA}) &= \tfrac{1}{2} \left[ 1 + \tfrac{\Delta}{U-V} \right] - \tfrac{1}{2} \sqrt{\left[1 + \tfrac{\Delta}{U-V}\right]^2 - 1} \le \tfrac{1}{2} ,\notag \\
 T(\mathrm{BRPA}) &= \tfrac{1}{4} \tfrac{U-V}{U + 2 t} \le \tfrac{1}{4} .
\end{align}
The bounds on $T$ are achieved at dissociation and for $\Delta = 0$ in the RPA case,
 which deviates from the BRPA value of $\Delta$.
The approximations change the fundamental range of $T$ values.

%PB.4 - Fitting parameters of H_2 model
We determine one set of parameters for the semiempirical model by solving
 for $\{t,\mu,V\}$ given $\{E_{\mathrm{HF}}, E_{\mathrm{MP2}}, E\}$ at $\theta = \tfrac{1}{4} \pi$,
\begin{align}
 E_1 &= 4(E_{\mathrm{HF}}-E_{\mathrm{MP2}})(E_{\mathrm{HF}}-E)/(E_{\mathrm{MP2}}-E), \notag \\
 V &= U + E_1 \left[ 1 - \sqrt{1 + (E - E_{\mathrm{HF}} + 2 U)/E_1} \right], \notag \\
 t &=  \tfrac{1}{16}(U-V)^2/(E_{\mathrm{HF}} - E) - \tfrac{1}{4}(E_{\mathrm{HF}} - E), \notag \\
 \mu &= t  - \tfrac{1}{4}(U+V) - \tfrac{1}{2}(E_0 - E_{\mathrm{HF}}).
\end{align}
 where $E_0 = 1/R$ for interatomic separation $R$ and $U$ is set to its value at the dissociation limit, $U = 0.5695$.
The H$_2$ data is generated in \textsc{gaussian09} using the cc-pV6Z basis \cite{G09}.
The data and parameters are compiled in Table \ref{H2data}.

%PB.5 - Pairwise extension to H_n
We consider an extension to H$_n$ having a pairwise form,
 \begin{align} \label{Hn}
 E_0 &= \frac{1}{2} \sum_{x \neq y} \frac{1}{R_{xy}} , \ \ \ x,y \in \{1, \cdots, n\}, \notag \\
 h_{(x \sigma )(y \sigma')} &= - \delta_{\sigma \sigma'} \delta_{xy} \tfrac{1}{2}  - \delta_{\sigma \sigma'} (1 - \delta_{xy}) t(R_{xy}) \notag \\
 & \ \ \ + \delta_{\sigma \sigma'} \delta_{xy} \sum_{z \neq x} \Delta \mu(R_{xz}) , \notag \\
 V_{(x \sigma)(y \sigma')} &= \delta_{xy} 0.5695  +  (1 - \delta_{xy}) V(R_{xy}) ,
\end{align}
 with spin indices, $\sigma \in \{ \uparrow, \downarrow \}$, and interatomic distances, $R_{xy}$.
Such a simple model will have limited transferability and care is required in fitting the parameter functions.

%PB.6 - Ring geometry, ring orbitals, ring HF eigenvalues/energy?
For simplicity, we restrict the scaling study to uniform H$_n$ rings that obey H\"{u}ckel's rule, $n = 4m + 2$.
Here, $R_{xy}$ is
\begin{equation}
 R_{xy} = R \sqrt{\left[ 1 - \cos\left(\tfrac{2 \pi (x-y)}{n}\right)\right] \Big/ \left[1 - \cos\left(\tfrac{2 \pi}{n}\right)\right]},
\end{equation}
 for nearest-neighbor distance $R$.
Spin and spatial symmetries simplify the normalized orbitals to
\begin{align}
 \phi_{(1\sigma)(x\sigma')} &= \delta_{\sigma\sigma'} \sqrt{\tfrac{1}{n}} , \ \ \
 \phi_{(n\sigma)(x\sigma')} = \delta_{\sigma\sigma'} \sqrt{\tfrac{1}{n}} (-1)^x , \notag \\
 \phi_{(2m\sigma)(x\sigma')} &= \delta_{\sigma\sigma'} \sqrt{\tfrac{2}{n}} \sin\left(\tfrac{2 \pi m x}{n}\right) ,  \ \ \ 1 \le m \le \tfrac{n}{2} - 1 , \notag \\
 \phi_{(2m+1\sigma)(x\sigma')} &= \delta_{\sigma\sigma'} \sqrt{\tfrac{2}{n}} \cos\left(\tfrac{2 \pi m x}{n}\right) ,
\end{align}
The density matrix sums to form the Dirichlet kernel,
\begin{equation}
 \rho_{(x \sigma )(y \sigma')} = \frac{1}{n} \delta_{\sigma\sigma'} \frac{\sin(\pi (x-y) / 2)}{\sin(\pi (x-y) / n)}.
\end{equation}
We use the HF energy of H$_6$, $E_6$, for fitting to these H$_n$ rings.
$R$ is optimized by minimizing the HF total energy.

%PB.7 - Final fit & its accuracy
We fit Eq. (\ref{Hn}) to $\{ E_{\mathrm{HF}}, E_6, E \}$ with exponential forms
 for $t(R)$, $\Delta \mu(R)$, and $V(R)$ plus long-range asymptotic corrections 
 for $\Delta \mu(R)$ and $V(R)$ with the form of an electrostatic potential between a 1s Slater-type orbital and point charge.
With 6 free parameters,
 the result of this nonlinear least-squares fit is
\begin{align} \label{pfunc}
 V(R) &=  0.304 \exp(-0.616 R) + [1-\exp(-0.616 R)]/R , \notag \\
 \Delta \mu(R) &= 0.000 \exp(-1.235 R) - [1 - \exp(-1.235 R)]/R , \notag \\
 t(R) &= 0.206 \exp(-0.383 R) ,
\end{align}
 with a root-mean-square error of $0.008$ Ha/atom.
The primed parameters calculated from Eq. (\ref{pfunc}) are given in Table \ref{H2data}.
In this model, H$_n$ rings have optimal $R$ values between $1.36$ and $1.89$
 and their HF energy asymptotes to $-0.527$ Ha/atom.

%T2
\begin{sidewaystable}[ph]
\caption{\label{H2data} Total energies (of H$_2$ unless otherwise stated) and model parameters (in Hartrees) at nearest-neighbor interatomic separation $R$ (in Bohrs). }
\begin{tabular}{r r r r r r r r r r r r r r r r}
\hline \hline
 $R$ & HF(H$_6$) & HF & B3LYP & UB3LYP & MP2 & exact & $\mu$ & $\mu'$ & $t$ & $t'$ & $U$ \& $U'$ & $V$ & $V'$ \\
\hline
$0.5$   &  $3.35175$ & $-0.48666$ & $-0.53368$ & $-0.53368$ & $-0.52150$ & $-0.52641$ & $-1.09542$ & $-1.42142$ & $0.31841$ & $0.17010$ & $0.56950$ & $0.11250$ & $0.75358$ \\
$0.7$   & $-0.27931$ & $-0.88259$ & $-0.92834$ & $-0.92834$ & $-0.91676$ & $-0.92184$ & $-1.01031$ & $-1.32677$ & $0.31680$ & $0.15756$ & $0.56950$ & $0.11662$ & $0.69790$ \\
$0.9$   & $-1.86432$ & $-1.04433$ & $-1.08972$ & $-1.08972$ & $-1.07811$ & $-1.08348$ & $-0.93969$ & $-1.24548$ & $0.31076$ & $0.14594$ & $0.56950$ & $0.12141$ & $0.64749$ \\
$1.1$   & $-2.62333$ & $-1.11045$ & $-1.15616$ & $-1.15616$ & $-1.14410$ & $-1.14991$ & $-0.88345$ & $-1.17541$ & $0.30041$ & $0.13518$ & $0.56950$ & $0.12688$ & $0.60180$ \\
$1.3$   & $-2.99665$ & $-1.13202$ & $-1.17854$ & $-1.17854$ & $-1.16578$ & $-1.17221$ & $-0.84003$ & $-1.11478$ & $0.28621$ & $0.12521$ & $0.56950$ & $0.13300$ & $0.56036$ \\
$1.5$   & $-3.17349$ & $-1.13137$ & $-1.17909$ & $-1.17909$ & $-1.16546$ & $-1.17273$ & $-0.80743$ & $-1.06210$ & $0.26887$ & $0.11598$ & $0.56950$ & $0.13964$ & $0.52271$ \\
$1.7$   & $-3.24462$ & $-1.11934$ & $-1.16858$ & $-1.16858$ & $-1.15399$ & $-1.16233$ & $-0.78364$ & $-1.01617$ & $0.24918$ & $0.10742$ & $0.56950$ & $0.14664$ & $0.48849$ \\
$1.9$   & $-3.25702$ & $-1.10160$ & $-1.15262$ & $-1.15262$ & $-1.13702$ & $-1.14673$ & $-0.76666$ & $-0.97595$ & $0.22810$ & $0.09950$ & $0.56950$ & $0.15375$ & $0.45734$ \\
$2.1$   & $-3.23641$ & $-1.08123$ & $-1.13425$ & $-1.13425$ & $-1.11764$ & $-1.12904$ & $-0.75476$ & $-0.94059$ & $0.20649$ & $0.09216$ & $0.56950$ & $0.16070$ & $0.42896$ \\
$2.3$   & $-3.19745$ & $-1.05996$ & $-1.11519$ & $-1.11519$ & $-1.09758$ & $-1.11105$ & $-0.74644$ & $-0.90939$ & $0.18512$ & $0.08537$ & $0.56950$ & $0.16727$ & $0.40307$ \\
$2.5$   & $-3.14877$ & $-1.03878$ & $-1.09640$ & $-1.09640$ & $-1.07784$ & $-1.09380$ & $-0.74049$ & $-0.88175$ & $0.16459$ & $0.07907$ & $0.56950$ & $0.17324$ & $0.37942$ \\
$3.0$   & $-3.01377$ & $-0.98932$ & $-1.05352$ & $-1.05444$ & $-1.03302$ & $-1.05718$ & $-0.73048$ & $-0.82513$ & $0.11939$ & $0.06529$ & $0.56950$ & $0.18470$ & $0.32871$ \\
$3.5$   & $-2.88444$ & $-0.94687$ & $-1.01831$ & $-1.02916$ & $-0.99677$ & $-1.03171$ & $-0.72172$ & $-0.78192$ & $0.08458$ & $0.05391$ & $0.56950$ & $0.19054$ & $0.28783$ \\
$4.0$   & $-2.77154$ & $-0.91161$ & $-0.99060$ & $-1.01666$ & $-0.96934$ & $-1.01625$ & $-0.71169$ & $-0.74821$ & $0.05929$ & $0.04452$ & $0.56950$ & $0.19123$ & $0.25460$ \\
$4.5$   & $-2.67665$ & $-0.88272$ & $-0.96928$ & $-1.01074$ & $-0.94992$ & $-1.00788$ & $-0.70042$ & $-0.72136$ & $0.04140$ & $0.03676$ & $0.56950$ & $0.18795$ & $0.22734$ \\
$5.0$   & $-2.59835$ & $-0.85919$ & $-0.95312$ & $-1.00788$ & $-0.93749$ & $-1.00369$ & $-0.68868$ & $-0.69958$ & $0.02880$ & $0.03035$ & $0.56950$ & $0.18205$ & $0.20478$ \\
$5.5$   & $-2.53436$ & $-0.84008$ & $-0.94096$ & $-1.00646$ & $-0.93098$ & $-1.00170$ & $-0.67710$ & $-0.68161$ & $0.01988$ & $0.02506$ & $0.56950$ & $0.17464$ & $0.18595$ \\
$6.0$   & $-2.48232$ & $-0.82454$ & $-0.93187$ & $-1.00573$ & $-0.92941$ & $-1.00078$ & $-0.66608$ & $-0.66657$ & $0.01352$ & $0.02069$ & $0.56950$ & $0.16651$ & $0.17008$ \\
$7.0$   & $-2.40582$ & $-0.80152$ & $-0.92003$ & $-1.00513$ & $-0.93770$ & $-1.00017$ & $-0.64641$ & $-0.64283$ & $0.00567$ & $0.01411$ & $0.56950$ & $0.15010$ & $0.14502$ \\
$8.0$   & $-2.35516$ & $-0.78604$ & $-0.91341$ & $-1.00496$ & $-0.95637$ & $-1.00004$ & $-0.63013$ & $-0.62499$ & $0.00157$ & $0.00962$ & $0.56950$ & $0.13524$ & $0.12630$ \\
$10.0$ & $-2.29692$ & $-0.76768$ & $-0.90729$ & $-1.00489$ & $-1.00903$ & $-1.00000$ & $-0.60596$ & $-0.60000$ & $-0.00176$ & $0.00447$ & $0.56950$ & $0.11195$ & $0.10043$ \\
$12.0$ & $-2.26638$ & $-0.75758$ & $-0.90475$ & $-1.00488$ & $-1.06870$ & $-1.00000$ & $-0.58946$ & $-0.58333$ & $-0.00272$ & $0.00208$ & $0.56950$ & $0.09564$ & $0.08347$ \\
$16.0$ & $-2.23446$ & $-0.74655$ & $-0.90242$ & $-1.00488$ & $-1.19169$ & $-1.00000$ & $-0.56858$ & $-0.56250$ & $-0.00300$ & $0.00045$ & $0.56950$ & $0.07473$ & $0.06251$ \\
$32.0$ & $-2.18483$ & $-0.73088$ & $-0.89929$ & $-1.00488$ & $-1.70376$ & $-1.00000$ & $-0.53588$ & $-0.53125$ & $-0.00231$ & $0.00000$ & $0.56950$ & $0.04056$ & $0.03125$ \\
$64.0$ & $-2.16530$ & $-0.72307$ & $-0.89772$ & $-1.00488$ & $-2.75324$ & $-1.00000$ & $-0.51848$ & $-0.51562$ & $-0.00142$ & $0.00000$ & $0.56950$ & $0.02134$ & $0.01563$ \\
$\infty$ & $-2.14577$ & $-0.71526$ & $-0.89616$ & $-1.00488$ & $-\infty$       & $-1.00000$ & $-0.50000$ & $-0.50000$ & $0.00000$ & $0.00000$ & $0.56950$ & $0.00000$ & $0.00000$ \\
\hline \hline
\end{tabular}
\end{sidewaystable}


\begin{thebibliography}{99}
\bibitem{symmetry_note}
In \textsc{inBRPAf}, tensors of the form $Z_f^e$ and $Z_{xy}^e$ have symmetries $Z_f^{\overline{e}} = (Z_{\overline{f}}^e)^*$
 and $Z_{xy}^{\overline{e}} = (Z_{xy}^e)^*$.
\textsc{MP2f} contains these symmetries and adds $Z_x^{\overline{e}\underline{p}} = \left(Z_x^{e\underline{p}^*}\right)^*$,
 $Z_{xy}^{\underline{p}^*0} = \left(Z_{xy}^{\underline{p}0}\right)^*$, and
 $Z_{\underline{a}\underline{i}}^{\overline{e}} = \left(Z_{\underline{a}^* \underline{i}^*}^e\right)^*$.
 \textsc{outBRPAf} contains these symmetries and adds
  $Z_x^{\underline{a}^* \underline{i}} = \left(Z_x^{\underline{a}\underline{i}^*}\right)^*$.
Only half the entries of these symmetric tensors need to be stored and calculated,
 which halves the memory and operation counts.
Complex arithmetic doubles memory usage and FFT runtimes
 and triples matrix multiplication runtimes over real arithmetic.
The additional cost of complex arithmetic is mitigated
 for all the operations except matrix multiplication (\textsc{zgemm}), 
 which is reduced to 1.5 times the cost model.
\bibitem{cost_note} The leading-order cost model assumes that all parameters are much larger than one, which is false for $\alpha = 2$.
We correct the conventional algorithm costs with
 $\alpha^3 \rightarrow \alpha (\alpha-1)^2$ for the operation counts,
 $3 \alpha^2 \rightarrow 3 \alpha^2 + 4 \alpha - 2$ for the MP2 memory cost, and
 $\alpha^2 \rightarrow \alpha (\alpha-1)$ for both BRPA memory costs.
\end{thebibliography}
\end{document}